\documentclass[a4paper,11pt,notitlepage]{article}
\usepackage[title,titletoc]{appendix}
\newcommand*\diff{\mathop{}\!\mathrm{d}}

\usepackage[margin=1in]{geometry}
\usepackage{apacite}
\usepackage{natbib}
\usepackage{csquotes}
\usepackage{import}
\usepackage{amssymb}

\usepackage{amsmath}
\usepackage{mathtools}
\usepackage{pifont}
\usepackage{bbm}
\usepackage{bm}
\usepackage{dsfont}
\usepackage[ruled,vlined]{algorithm2e}

\usepackage{authblk}
\usepackage{hyperref}
\usepackage{graphicx}
\usepackage[title,titletoc]{appendix}

\usepackage{wrapfig}
\usepackage{titling}
\usepackage{float}
\usepackage{tikz}

\usepackage{caption}
\usepackage{subcaption}

\DeclareMathOperator*{\argmin}{argmin}

\usepackage{pdfpages}

\title{A framework for statistical modelling of the extremes of longitudinal data, applied to elite swimming}
\author{Jess Spearing$^1$, Jonathan Tawn$^1$, David Irons$^2$, Tim Paulden$^2$\\
$^1$Lancaster University, $^2$ATASS Sports}
\begin{document}

\maketitle
\begin{abstract}
We develop methods, based on extreme value theory, for analysing observations in the tails of longitudinal data, i.e., a data set consisting of a large number of short time series, which are typically irregularly and non-simultaneously sampled, yet have some commonality in the structure of each series and exhibit independence between time series. Extreme value theory has not been considered previously for the unique features of longitudinal data. Across time series the data are assumed to follow a common generalised Pareto distribution, above a high threshold. To account for temporal dependence of such data we require a model to describe (i) the variation between the different time series properties, (ii) the changes in distribution over time, and (iii) the temporal dependence within each series. Our methodology has the flexibility to capture both asymptotic dependence and asymptotic independence, with this characteristic determined by the data.
Bayesian inference is used given the need for inference of parameters that are unique to each time series. Our novel methodology is illustrated through the analysis of data from elite swimmers in the men's 100m breaststroke. Unlike previous analyses of personal-best data in this event, we are able to make inference about the careers of individual swimmers - such as the probability an individual will break the world record or swim the fastest time next year.
\end{abstract}

\section{Introduction}
\label{sec:introduction}

Traditional statistical techniques are designed to describe the behaviour of the ``typical" data and many analyses involve the identification and removal of observations from the tails of the data to improve robustness. But what if the data of most interest \textit{are} those observations in the tails? When considering natural disasters such as flooding, stresses or corrosion on a structure, financial crises, or sporting records, it is precisely these \textit{extreme} values that are most pertinent. \textit{Extreme value theory} (EVT) is a branch of statistics specifically designed to model such extreme or rare events, with the methods having a strong probabilistic framework based on asymptotic justifications. This paper presents novel methodology for the analysis of longitudinal data where the extreme values are of primary interest.

Early EVT methods describe the extremal behaviour of independent univariate random variables, possibly in the presence of covariates, with the book of \cite{coles2001introduction} an accessible introduction. Since then, the extremal properties of ever more rich data structures have been studied.
For univariate stationary processes the following features have been considered: long- and short-range dependence \citep{ledford2003diagnostics}, Markov structure \citep{winter2017k}, and hierarchical clustered data \citep{smith2000bayesian,dupuis2023modeling, momoki2023mixed}.
For multivariate extreme value problems, 
structure has been identified and exploited through the use of graphical structures \citep{engelke2020graphical}
and models for conditional structures through asymptotic independence \citep{heffernan2004conditional}.  Various approaches have also been developed for spatial, and spatial temporal extreme events, such as $r$-Pareto processes \citep{de2022functional}, spatial conditional asymptotically independent processes \citep{wadsworth2022higher}, 
and for spatial mixture processes \citep{richards2023joint}.

Currently there is no EVT methodology to model longitudinal (or panel) data. Such data comprises a number of {\it subjects}, with each subject recording a time series of responses
\citep{diggle2002analysis}. 
Specifically, there are a set of subjects, $\mathcal{I}$, with a subject $i$ having responses $\mathcal{J}_i$, for all $i\in\mathcal{I}$. The response $X_{i,j}$ belonging to subject $i$, occurs at time $t_{i,j}\in\mathbb{R}$, for all $j\in\mathcal{J}_i,\;i\in\mathcal{I}$. The typical assumptions made about the collection $\{X_{i,j}: 
 j\in \mathcal{J}_i, \mbox{ for }i\in\mathcal{I}\}$ are that: the $X_{i,j}$ are independent over different $i \in\mathcal{I}$, irrespective of $j$, but they are potentially dependent across $j\in \mathcal{J}_i$ for any given $i\in\mathcal{I}$; 
 there are a large number of subjects relative to the number of responses per subject; 
 and the distribution of $X_{i,j}$ varies with $t_{i,j}$ similarly over subjects.

 For analysing the extremes of longitudinal data, the \textit{sample} $\mathcal{I}$ comprises those subjects with at least one extreme observation within the observed time-frame.
We distinguish between this sample of subjects $\mathcal{I}$, and the \textit{population} of extreme subjects, which includes those subjects with extreme responses that are exclusively outside the observed time-frame; i.e.,  the subjects may have either no responses at all, or have responses that are exclusively non-extreme.
In applications where subjects exhibit non-stationarity, future extreme events change from being from subjects in $\mathcal{I}$ to responses on subjects in the broader population.
 
Longitudinal data analyses arise most commonly in designed trials (e.g., in clinical or corrosion contexts) whereby multiple subjects (e.g., patients or material coupon samples) have a single quantity (e.g., blood pressure or corrosion, respectively) measured over time. There has been  no extreme value modelling of clinical and corrosion data which captures the full specification of such data. For example, 
\cite{southworth2012extreme} and \cite{laycock1993exceedances} do not consider repeated measurements  on the same subject. \cite{fougeres2006pitting} do consider multiple observations per coupon but assume that observations from the same coupon are IID. 
Further differences between our approach and papers which model extremes of longitudinal/panel data are outlined  in the supplementary material.
Our paper aims to be the first foray into developing broadly usable EVT methods for longitudinal data, with the flexibly to model both asymptotic dependent and asymptotic independent temporal extremal dependence structures and to capture trends in the means of subjects' responses over time.

Extreme value analysis of longitudinal data is important in athletics and swimming, with clear relevance for studying the progression of records and predicting who will be fastest next year. Athletes/swimmers (subjects) all strive to be fastest in their event, with their personal career progression  having stages of improvement and decline with age, and with them competing at irregular and non-synchronised times. These subject-specific trends arise whilst overall performances by the elite athletes/swimmers are improving over time.

The application of EVT methods is not new for sports' data. EVT is used by \cite{stephenson2013determining} to model athletics times data and by \cite{strand1998modeling} to estimate the peak age of competitive 10K road race runners. 
\cite{spearing2021ranking} use EVT to model the evolution of elite swimming over time, including the effect of different swim-suit technologies, and combine data across different swimming strokes, gender categories and distances through the use of a data-based covariate. 
These models do not attempt to model dependence structure - either they assume that performances from the same subject are independent of each other, or only incorporate each subject's best performance into the data set.
Each approach leads to incomplete inference: the former produces an underestimation of standard errors and confidence interval widths when the independence assumptions are invalidated; and the latter uses a smaller data set than is available, leading to inefficient inference. However, the true limitation of these simplifications runs deeper. The lack of any longitudinal structure in these models means that no statistical inference can be conducted on any facet involving individual competitors.
 
We illustrate our novel EVT methodology for longitudinal data in the context of elite swimming, for the mens' 100m breaststroke (long course) event. A swimmer is defined as elite if they have ever produced a swim-time less than a certain threshold $u$. The selection of this threshold $u$, discussed in the supplementary material, is here
taken as the 200th fastest personal-best swim-time in the mens' 100m breaststroke event, which is $u=61.125$ seconds. In our approach (i) all the available recorded swims from each elite swimmer are modelled, irrespective of whether they are below or above $u$, (ii) the swimmer who produced each swim-time is accounted for, as is their age at which it was achieved, (iii) the dependence between swim-times from the same swimmer is captured, with this dependence allowed to weaken as the inter-swim-time increases.

\begin{figure}[h!]
    \centering
    \includegraphics[width=0.5\textwidth]{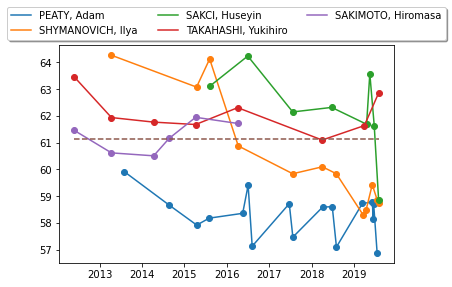}
    \caption{Data for swim-times (in seconds) plotted against the date when it was achieved for the mens' 100m breaststroke (long course) event. All competition best performances  are shown for five swimmers over time. The dashed line indicates the threshold $u$.}
    \label{fig:swimmerdata}
\end{figure}

Figure~\ref{fig:swimmerdata} depicts the competition-best swim-times for five of the 200 elite swimmers who epitomise the range of career trajectories.
Of these swimmers, Adam Peaty holds the current world record and so, the fastest personal-best (PB). Ilya Shymanovich has the 2nd fastest PB in the data, Sakci Hueseyin 8th, and Sakimoto Hiromasa the 101st.  Takahashi has the 196th fastest PB, which is only just faster than $u$ with that being their only swim faster than $u$.
The performances, and \textit{career trajectories}
of the top two swimmers differ. Peaty is consistently fast, producing the seven fastest times of the competition-best dataset, and with all his performances faster than $u$. Conversely, Shymanovich is in a clear progression stage of his career, moving from being slower than $u$ to consistently faster. 
The figure illustrates there to be differing strategies for which, and how many, competitions swimmers compete in.

Now consider the marginal distribution of the extreme swimming values, i.e., the values below $u=61.125$. To motivate a possible model for these values we draw on EVT which provides an asymptotic justification for using the generalised Pareto distribution (GPD), 
however, we have no justifiable parametric model for observations slower than $u$. In modelling the extremes of longitudinal data, it is desirable that the extreme data be the most influential. Therefore, observations slower than $u$ are treated as censored at the level of the threshold. 
As a consequence, all but one of Takahashi's observations are censored, whereas all of Peaty's observations can be modelled with the GPD. Critically, the values slower than the threshold are not lost as they provide marginal information about the rate of performing better than $u$ and they inform about the dependence structure for individual swimmers through information about patterns of better and worse performances relative to $u$.

Conventional presentation of EVT pertains to the largest values - or equivalently the upper tail, yet the best swim-times are the smallest - or in the lower tail. By applying our methodology to \textit{negative} swim-times, standard EVT results  can be utilised. So, throughout we present theory and methods for the upper extremes of longitudinal  data.
Section~\ref{sec:Theory} presents the
extensions of univariate EVT to cover the time series aspect of each subject's data and illustrates how the level of subject variation induces both {\it asymptotic dependence} and {\it asymptotic independence}.
Section~\ref{sec:modelling} contains the main contribution of the paper - a novel approach to the modelling of the extremes of longitudinal data. 
Section~\ref{sec: inference} presents the general Bayesian inference framework and 
Section~\ref{sec: application} details how this modelling and inference framework can be applied to the elite swimming data, and provides examples of particular inferences and predictions that are available using our methodology. A discussion and future work is in Section~\ref{sec: discussion}.

\section{Motivating Theory}
 \label{sec:Theory}
\subsection{Univariate extremes}
\label{sec:Univariate}
In its simplest form, univariate extreme value theory (EVT) applies to independent and identically distributed (IID) random samples $Y_1,\dots, Y_n$, where each variable has continuous distribution function $F$. The block maxima and peaks over threshold methods are the two core approaches in univariate EVT \citep{coles2001introduction}. We are interested in formulating a theoretically justified marginal extreme value model for temporally dependent variables and describing the dependence structure induced by within- and across-subject observations for longitudinal data. 
We also consider a stationary process, 
$X_1, \ldots ,X_n$ which also has the marginal distribution function $F$ but satisfies conditions such that its long-range dependence is restricted to behave as effectively independent, see 
\cite{leadbetter2012extremes} for their precise form and discussion of the limit results \eqref{eq:probMax} and \eqref{eq:probMaxDep}.
 Under such conditions, the following results hold. If $M_{Y,n}:= \max\{Y_1,\dots, Y_n\}$
 and there exist norming sequences $a_n > 0$ and $b_n$, such that
\begin{equation}
\Pr\left\{\frac{M_{Y,n}-b_n}{a_n} \leq x \right\} = F^n(a_n x + b_n) \rightarrow G(x), \text{ as } n\rightarrow \infty,
\label{eq:probMax}
\end{equation}
where that the limiting distribution $G(x)$ is non-degenerate, then 
$G(x)$ must be a generalised extreme value (GEV) distribution, which has the form
$G(x) = \exp\left(-[1+\xi(x-\mu)/\sigma]^{-1/\xi}_+\right)$,
where $\mu,\; \xi \in \mathbb{R}, \;\sigma \in \mathbb{R}^+$, are the location, shape and scale parameters respectively and with the notation $ y_+ := \max(y,0)$. Then for $M_{X,n}:= \max\{X_1,\dots, X_n\}$, if $(M_{X,n}-b_n)/a_n$ has a non-degenerate limit distribution, as $n\rightarrow \infty$, it follows that
\begin{equation}
\Pr\left\{\frac{M_{X,n}-b_n}{a_n} \leq x \right\}  \rightarrow [G(x)]^{\theta}, \text{ as } n\rightarrow \infty,
\label{eq:probMaxDep}
\end{equation}
where $0<\theta\le 1$ is the extremal index; a measure of extremal temporal dependence.

We are primarily interested in having an asymptotically motivated model for the upper tail behaviour of $\{X_t\}$ and $\{Y_t\}$. These models are derived directly from the limiting distribution of block maxima identified above. First, denote $D_G:=\{x\in \mathbb{R}: 0<G(x)<1\}$ and let both $x$ and $u$ be in $D_G$ with $x>u$. Then, 
as $n \rightarrow\infty$, applying a Taylor series approximation to limit \eqref{eq:probMax} gives,
$n[1-F(a_n x + b_n)] \rightarrow -\log G(x) = [1+\xi(x-\mu)/\sigma]^{-1/\xi}_+$ and for $Y\sim F$,
\begin{equation}
\Pr\{Y > a_n x + b_n | Y > a_n u + b_n\} \rightarrow \log G(x)/ \log G(u) =: \bar{H}_u(x),
\label{eq:probExc}
\end{equation}
with $\bar{H}_u(x) := 1-H_u(x)$, and where the distribution function $H_u$ is given by
\begin{equation}
H_u(x) = 1-\left[1+\xi \left(\frac{x-u}{\sigma_u}\right)\right]^{-\frac{1}{\xi}}_+.
\label{eq:GPd}
\end{equation}
where $\sigma_u = \sigma + \xi(u-\mu)$. The distribution function $H_u$ is termed the generalised Pareto distribution (GPD), denoted GPD$(\sigma_u,\xi)$,
with threshold $u$, shape parameter $\xi \in\mathbb{R}$ and scale parameter $\sigma_u \in \mathbb{R}_+$. 
For $\xi < 0$, there exists a finite value $x^H = u - \sigma_u/\xi: \; H_u(x) = 1, \; \forall x>x^H $, whereas for $\xi\geq 0, \; H_u(x) < 1, \; \forall x<\infty$. This GPD result is powerful as it holds as the limit distribution for a very broad class of continuous distributions $F$.

The same GPD$(\sigma_u,\xi)$ limit distribution holds for $\Pr\{X > a_n x + b_n | X > a_n u + b_n\}$ as $n\rightarrow \infty$ with $X\sim X_i$. 
Additionally \citet{leadbetter1991basis}  gives that for an arbitrary cluster maxima $X_C$ of $\{X_t\}$, then 
$\Pr\{X_C > a_n x + b_n | X_C > a_n u + b_n\}$ as $n\rightarrow \infty$, is also GPD$(\sigma_u,\xi)$. This has motivated the use of the generalized Pareto distribution as a
statistical model for cluster maxima
\citep{davison1990models}, but  for our purposes shows the connection between the tail of the distribution for all swims and competition maxima. 

In practice the limit distribution~\eqref{eq:probExc} is assumed to hold exactly for some finite $n$, or equivalently for some fixed threshold $a_nu+b_n$, corresponding to a high quantile of $Y$ or $X$. A consequence  is that the limit distribution $H_u$ gives an asymptotic model, 
determined by only two parameters, 
for the distribution of exceedances above a threshold $u$, no matter the form of marginal distribution $F$. 
To complete the description of the tail of the marginal distribution we define the marginal probability of an threshold exceedance, $\lambda_{u}:=\Pr(X>u)$. 
The optimal choice of $u$ is determined by bias-variance trade-off arguments \citep{scarrott2012review}.

\subsection{Extremal dependence: measures and modelling strategies}
\label{sec:AsymptoticDependence}
To account for dependence between the extreme responses from a given subject, we draw on knowledge of generic extremal  dependence measures and the associated modelling strategies before considering the specific features that are unique to longitudinal data.

When modelling dependence between the extremes of two variables the typical  approach involves first deciding on the \textit{form} of extremal dependence, and then looking for an appropriate model formulation subject to that form \citep{coles1999dependence}.
For bivariate extremes, with continuous random variables 
$(X_1,X_2)$ with marginal distributions $F_1$ and $F_2$, respectively,
the two forms of extremal dependence in the upper tail are determined by {\it the coefficient of asymptotic dependence} $\chi := \lim_{q\uparrow 1} \chi(q)$ where,
for $0<q<1$,
\begin{equation}
    \label{eq:extremal dependence}
   \chi(q) := \Pr\{F_1(X_1)>q~|~F_2(X_2)>q\}=
\Pr\{F_1(X_1)>q,F_2(X_2)>q\}/(1-q),
\end{equation}
with {\it asymptotic dependence} given by $0<\chi\le 1$ and 
{\it asymptotic independence} by $\chi=0$. In essence, asymptotic dependence allows the very largest values of $X_1$ and $X_2$ to occur together, unlike for asymptotic independence. This interpretation is made precise by looking at the limiting distribution of normalised componentwise maxima of IID vectors $\{(X_{1i},X_{2i}): i=1, \ldots ,n\}$, such that the marginal limiting distributions are non-degenerate. Then, the two variables are termed asymptotic dependent, or asymptotic independent, if that limiting distribution exhibits dependence, or independence, respectively. Variables may exhibit extremal dependence without asymptotic dependence, with this dependence measured by the {\it coefficient of asymptotic  independence}, $\bar{\chi}:= \lim_{q\uparrow 1}\bar{\chi}(q) \in (-1,1]$, where for $0<q<1$,
\begin{equation}
    \label{eq: coefficient asymptotic independence}
    \bar{\chi}(q):= \frac{2\log \Pr\{F_2(X_2)>q\}}{\log \Pr\{F_1(X_1)>q,F_2(X_2)>q\}}-1,
\end{equation}
 with independent variables giving $\bar{\chi}=0$, and $0<\bar{\chi}<1$ $(\bar{\chi}<0)$ corresponding to a positive (negative) extremal dependence form of asymptotic independence respectively, and $\bar{\chi}=1$ under asymptotically dependence. Both $\chi$ and $\bar{\chi}$ are invariant to the marginal
distributions, so in terms of models for the joint distribution it is helpful to consider different copulas \citep{nelsen2007introduction}. 

\cite{fougeres2009models} use the bivariate extreme value distribution copula  with logistic$(\alpha)$ dependence structure,  
which has $(\chi,\bar{\chi})=(2-2^{\alpha},1)$ 
for $0\le \alpha<1$ and $(\chi,\bar{\chi})=(0,0)$ when $\alpha=1$.
This copula is restrictive as it cannot capture positive dependence within the asymptotic independence case.
The Gaussian copula has  $(\chi,\bar{\chi})=(0,\rho)$ for correlation parameter $-1<\rho<1$ \citep{coles1999dependence}, though not offering asymptotic dependence, gives flexibility and parsimony of asymptotic independence structures  and it benefits from closed form conditional distributions for simulating the time series features of longitudinal data. 

Given these properties, within-subject measurements were modelled via a Gaussian copula, see Section~\ref{sec:dep_modelling}. This may appear restrictive, but we demonstrate in 
Section~\ref{sub:measures} that, due to the variation across subjects, any level of asymptotic dependence or asymptotic independence can be approximated for the longitudinal data using this copula. This flexibility is not possible if starting with an asymptotically dependent copula.

\subsection{Measures of longitudinal data extremal dependence}
\label{sub:measures} 
Consider a special case of the set up of Section~\ref{sec:introduction},
with a stationary continuous time process for each subject $i\in\mathcal{I}$ being $\{X_{i}(t)\}$ for all $t$ which are observed at a set of identical and equally spaced time points across the  $n$ subjects. Denote $X_{i,j}=X_{i}(t_{i,j})=X_{i}(t_{j})$, where $t_j$ is the $j$th time point. We assume that the marginal distribution of the $i$th subject is $F_i(\cdot)=F(\cdot;\alpha_i)$ where $F$ is a common continuous distribution function family  with parameter $\alpha_i\in \mathbb{R}$ which varies over $i\in\mathcal{I}$. We term $\alpha_i$ the \textit{attribute} of subject $i$, with the property that $F(x;\alpha_i)>F(x;\alpha_j)$ for all $x\in \mathbb{R}$ for all $\alpha_i>\alpha_j$. Increasing the attribute of a subject makes the quantiles of its response distribution larger. 
Given the potential heterogeneity between subjects, a basic application of the coefficient of asymptotic dependence for within-subject dependence
at time-lag $\tau$, for all $\tau \in \mathbb{R}$ for each subject $i\in \mathcal{I}$ is:
\begin{equation}
    \label{eq:AD measure - chi_i(tau)}
\chi_{i}(\tau):=\lim_{q\uparrow 1}\Pr(F(X_{i}(\tau);\alpha_i)>q \mid F(X_{i}(0); \alpha_i)>q),
\end{equation}
or the equivalent asymptotic independence measure $\bar{\chi}_{i}(\tau)$. These measures do not provide a global description of the dependence across all subjects in
$\mathcal{I}$, with two such measures being discussed in the supplementary material.

To study how subject attributes determine extremal dependence of longitudinal data,
consider all $n$ independent subjects having responses at only two time points - which are the same across subjects -  and the responses per subject are independent,  except for subject $n$. Additionally all subjects have identical attributes except for subject $n$. In the notation of Section~\ref{sec:introduction}, $\mathcal{J}_i=\{1,2\}$ for all $i\in\mathcal{I}$,   $X_{i,j}\sim N(0,1)$ for $i=1,\ldots ,n-1$ and $j=1,2$ are mutually independent, while subject $n$ has a potentially different mean, namely
$X_{nj}\sim N(\alpha_n,1)$ for $j=1,2$ and $(X_{n1},X_{n2})$ are bivariate Normal with correlation $0\le \rho<1$, which with standard margins has joint distribution function denoted by $\Phi_2(\cdot,\cdot;\rho)$. Thus here $F(x;\alpha_i)=\Phi(x-\alpha_i)$, with attributes $\alpha_1=\ldots =\alpha_{n-1}=0$ and $\alpha_n$. 

The subject-specific dependence measures at lag $\tau=1$, are $(\chi_{i1}, \bar{\chi}_{i1})=(0,0)$ for subjects $i=1, \ldots ,n-1$ due to the independence assumption, and 
 due to the bivariate Normal distribution for subject $n$ we have $(\chi_{i1}, \bar{\chi}_{i1})=(0,\rho)$. So there is asymptotic independence across  subjects, although subject $n$ is not independent. When studying the across population behaviour, we investigate two cases for $\alpha_n$ (i) $(2\log n)^{1/2}/\alpha_n=o(1)$ as $n\rightarrow \infty$ and (ii) $\alpha_n/(2\log n)^{1/2}=o(1)$ as $n\rightarrow \infty$, i.e., the latter includes both $\alpha_n\rightarrow \infty$ as $n\rightarrow \infty$ and $\alpha_n=0$ for all $n$. We will show that cases (i) and (ii) lead to results which are consistent with asymptotic independence and asymptotic dependence respectively.

Consider the dependence of the componentwise maxima $(M_{n,1},M_{n,2})$, over the two time points, i.e., 
$M_{n,j}:=\max\left(\{X_{i,j}:i\in \mathcal{I}\}\right)$, for $j=1,2$ and  
$n\rightarrow \infty$ for case (i). 
For the two marginal  maxima  we have that, for any $x\in\mathbb{R}$,  
$\Pr\{M_{nj}-\alpha_n<x\}=\left[\Phi(\alpha_n+x)\right]^{n-1}\Phi(x)
    \rightarrow   \Phi(x)$
as $n\rightarrow \infty$, i.e., a non-degenerate Gaussian limit. This result follows from Section~\ref{sec:Univariate} since for $\alpha_n$ in case (i), $n[1-\Phi(\alpha_n+x )]\rightarrow 0$ for all 
$x\in\mathbb{R}$. The reason for this convergence follows from univariate extreme value results for standard Gaussian variables, i.e., $n[1-\Phi(a_ny+b_n)]\rightarrow \exp(-y)$
for $a_n=(2\log n)^{-1/2}$ and $b_n=(2\log n)^{1/2}+o(1)$ for $y\in \mathbb{R}$
\citep{leadbetter2012extremes}. 
Now consider the joint probability, for $(x,y)\in\mathbb{R}^2$, as $n\rightarrow \infty$, given by 
\begin{equation}
    \label{eq:deplimitNormal}
    \Pr\{M_{n1}-\alpha_n<x, M_{n2}-\alpha_n<y\}=
    \left[\Phi(\alpha_n+x)\Phi(\alpha_n+y)\right]^{n-1}   \Phi_2(x,y; \rho)
   \rightarrow  \Phi_2(x,y; \rho), 
   \end{equation}
 where the non-degenerate limit arises using the same logic as for the marginal convergence. The joint maxima are asymptotically dependent when $\rho>0$, with the limit not restricted to being a bivariate extreme value distribution as the variables are not identically distributed. Case (ii) for the $\alpha_n$ gives that 
$\Pr\{(M_{nj}-b_n)/a_n<x\}\rightarrow G(x)$, where $G(x)=\exp[-\exp(-x)]$,
and   
\begin{align*}
\Pr\{(M_{n1}-b_n)/a_n<x, (M_{n2}-b_n)/a_n<y\}&\rightarrow 
G(x)G(y)
\end{align*}
as $n\rightarrow \infty$.
These limits show a change in the marginal limit distribution from Gaussian to Gumbel and independence of the limiting componentwise maxima, so asymptotic independence. 

These two asymptotic regimes for longitudinal data illustrate that the nature of extremal dependence is  different for this framework than for stationary series. Specifically, they demonstrate that asymptotic dependence per subject is not essential to achieve asymptotic dependence for longitudinal data; 
asymptotic dependence can be achieved by having subjects with a heavy tailed attribute distribution; and that both asymptotic dependence and asymptotic independence can be achieved from a simple Gaussian copula. Critical to the form of extremal dependence 
is the level of between-subject variation (via the attribute variation) relative to the within-subject variation.
Here in case (i) $\alpha_n$ dominates the maximum of the responses over all other subjects but not in case (ii).

\section{Extremal Model for Longitudinal Data}
\label{sec:modelling}
\subsection{Population Marginal Model}
\label{sub:modelling:subsec:popmodel}

When developing a marginal model for the population of longitudinal random variables $\{(X_{i,j}, t_{i,j}): j\in\mathcal{J}_i, i\in\mathcal{I}\}$, 
we make a critical decision of ignoring the subject-specific nature of the data as is conventional in previous extremal analyses. We refer to this characteristic as \textit{subject-ignorant}.
Instead, the information regarding specific subjects is captured through our dependence modelling in Section~\ref{sec:dep_modelling}. The reasons for this strategy are three-fold. Firstly, the number of observations per subject, e.g., $\mid \mathcal{J}_i\mid $ for subject $i$, is likely to be small in most applications and so a separate marginal model (see Section~\ref{sec:Univariate}) per subject for the data in the tails is an unrealistic target, even with some pooling \citep{dupuis2023modeling}.
Secondly, modelling the tail of a population using a single GPD enables inference to be made about trends in the population as a whole \citep{spearing2021ranking}.
Thirdly, this enables application specific structure identified from previous GPD analyses, which ignore subject knowledge, to be exploited.

Given the above strategy, consider a generic pair $(X,t)$, written as $X_t$.
For a selected constant over time threshold $u$, 
there are three features of the distribution of $X_t$ we describe: the behaviour above the threshold $u$, the probability of $X_t$ exceeding $u$, and the distribution of $X_t$ being below $u$. The latter is not typically studied in extremes of a univariate variable, but keeping track of the behaviour below the threshold is important here for dependence modelling of within-subject data in Section~\ref{sec:dep_modelling}.

Above the threshold $u$ we assume that for $x>0$, $\Pr\{X_t-u<x|X_t>u\}$
has a GPD$(\sigma_u(t), \xi)$, as given by expression~\eqref{eq:GPd}.
Although $X_t$ is potentially complex in its variation over $t$, temporal variation is assumed only through $\sigma_u$, a typical and pragmatic approach \citep{coles2001introduction}.
The probability of exceeding the threshold $\Pr\{X_t>u\} =: \lambda_u(t)$ is also allowed to vary with time. 
Literature on modelling approaches for how
$(\sigma_u(t),\lambda_u(t))$ vary with $t$ include parametric, see Section~\ref{sec: modelling swimming}, non-parametric, or machine learning approaches, see \citep{richards2022unifying}.

The $X_t$, conditionally on being below $u$, are assumed to follow some unknown but continuous density function $h_t:(\infty, u]\rightarrow \mathbb{R}_+$, with $\int_{-\infty}^u h_t(s) \diff s = 1$, where $h_t$ does not depend on $(\lambda_u, \sigma_u, \xi)$. Combining all these models gives the distribution function $F_{X_t}$ of $X_t$ as 
\begin{equation}
\label{eq:GPD marginal_cdf}
F_{X_t}(x) = \left\{\begin{matrix*}[l]
1-\lambda_u(t)\left[1 + \xi (x-u)/\sigma_u(t)\right]_+^{-\frac{1}{\xi}},&\quad x>u,\\
[1-\lambda_u(t)]\int_{-\infty}^x h_t(s) \diff s,&\quad x\leq u.
\end{matrix*}\right.
\end{equation}
As with the vast majority of extreme value modelling we avoid imposing a structure on the distribution of $X_t<u$, i.e., the density $h_t$ here. Even if a parametric model for $h_t$ had no parameters in common with those in the GPD or $\lambda_u$ models, there is a risk of bias from mis-specifying $h_t$ in the longitudinal setting due to the dependence between values $X_{i,j}$ and $X_{ij^{\prime}}$ for $j^{\prime} \not= j$, where $X_{i,j}<u<X_{ij^{\prime}}$. In such cases, errors in modelling below the threshold can induce errors above the threshold to compensate. Therefore,
any actual value $X_{i,j}$ below $u$ is instead treated as censored, i.e., as a realisation of the event $X_{i,j}<u$.

\subsection{Dependence Structure in a Latent Space}
\label{sec:dep_modelling}

The focus now turns to modelling the dependence structure of random variables $\{(X_{i,j}, t_{i,j}): j\in\mathcal{J}_i, i\in\mathcal{I}\}$.
Specifically, we need to allow for temporal dependence between within-subject variables and independence between across-subject variables, so unlike in Section~\ref{sub:modelling:subsec:popmodel} knowledge of each subject's contribution to the data is accounted for. The formulation of these models builds on the findings of Section~\ref{sub:measures}, which showed that multivariate Gaussian distributions for within-subject variations combined with an attribute distribution that has the capacity for both heavier and shorter tails than the within-subject Gaussian distribution, provide sufficient flexibility to allow for both extremal dependence forms.

The adopted modelling strategy bears likeness to that of \cite{huser2019modeling}, i.e., focusing on the joint structure of variables, without concern for its implications on the marginals at that stage. Subsequently, in Section~\ref{sec:transform}, the marginal distributions of this model are linked to the formulation in Section~\ref{sub:modelling:subsec:popmodel}.
In particular, a model is adopted in terms of variables $\{(Z_{i,j}, t_{i,j}): j\in\mathcal{J}_i, i\in\mathcal{I}\}$, where $Z_{i,j}=T_t(X_{i,j})$ for a function $T_t$ defined in Section~\ref{sec:transform}, and we refer to the stochastic model for the $\{Z_{i,j}\}$ as a model in the \textit{latent space}.

In the latent space we develop a model for responses from the same subject, e.g., $\{(Z_{i,j}, t_{i,j}): j\in\mathcal{J}_i\}$ for subject $i$.
We follow standard Gaussian modelling assumptions of longitudinal data analysis \citep{diggle2002analysis}.
The subject-specific model takes $Z_{i,j}$, across $j\in\mathcal{J}_i$, as realisations of a Gaussian process $Z_{i}(t)$ over time $t\in\mathbb{R}$ observed at the times $\pmb{t}_i:=\{t_{i,j}; j\in \mathcal{I}_i\}$. Specifically,  
\begin{equation}
Z_{i}(t) \sim \mathcal{GP}\left(\mu_i(t),\nu_i^2 K_{\pmb{\kappa}}(\cdot,\cdot)\right),\mbox{ for all }t\in \mathbb{R},
\label{eqn:GP}
\end{equation}
where the \textit{mean function} $\mu_i(t):\mathbb{R} \rightarrow\mathbb{R}$ is a subject-specific time-dependent mean, $\nu_i>0$ is a homogeneous subject-specific standard deviation, and 
$K_{\pmb{\kappa}}$ is a stationary kernel, which is shared over subjects, and which dictates the {\it subject-conditional correlation} between the process at any times $t\in \mathbb{R}$ and $t^\prime\in \mathbb{R}$ with hyper-parameters $\pmb{\kappa}$.
The term $\mu_i(t)$ allows for the statistical properties of individual subjects to evolve over time separately from that of the population marginal model, as is the case for many applications in longitudinal analysis.
To avoid over-parametrisation over individuals it is reasonable to assume that 
\begin{equation}
\mu_i(t;\pmb{\theta}_i, \pmb{\gamma})= \alpha_i+ \mu(t,\tau_i; \pmb{\gamma}), \mbox{ for all }t\in \mathbb{R},
\label{eqn:measurefunction}
\end{equation}
for a subject-ignorant function $\mu\le 0$ with parameters $\pmb{\gamma}$, subject-specific parameters $\pmb{\theta}_i=(\alpha_i, \tau_i)$ and covariates (which are ignored in this formulation, but are used in Section~\ref{sec: modelling swimming}).
To ensure that $\alpha_i$ is identifiable, the maximum of the function $\mu$, over $t$,  is set to zero, i.e., $\alpha_i=\max_{t\in\mathbb{R}}\mu_i(t; \pmb{\theta}_i)$. Then $\alpha_i$ is the $i$th subject's \textit{attribute}, as in Section~\ref{sub:measures}.
When $\mu \equiv 0$ in model~\eqref{eqn:GP}
the subject-specific dependence measures are 
$(\chi_{i\tau}, \bar{\chi}_{i\tau})=(0,K_{\pmb{\kappa}}(0,\tau))$, for all $i\in \mathcal{I}$.

The form of the stationary kernel is application specific. A powered exponential is used
\begin{equation}
\label{eq:kernel (exponential)}
K_{\pmb{\kappa}}(t, t^\prime) = \exp(-\kappa_0|t-t^\prime|^{\kappa_1}), 
\end{equation}
with $\pmb{\kappa}=(\kappa_0, \kappa_1) \in\mathbb{R}_+\times [0.5,2]$ in Section~\ref{sec: application},
where smaller $\kappa_0$ gives less subject-conditional dependence (with the limit $\kappa_0 \rightarrow \infty$ giving subject-conditional independence); and $\kappa_1$ influences the local smoothness of the process, with larger $\kappa_1$ giving a smoother process, with the limit $\kappa_1\rightarrow 2$
corresponding to a process which is infinity differentiable, and when $\kappa_1=1$ the process is Markov.
Other well-established kernels, e.g., the Mat\'ern family \citep{diggle2002analysis}, were trialled in exploratory analysis for the application in Section~\ref{sec: application} but made no practical differences due to having few observations per subject and none at short time lags.

Conditioning on the latent model parameters, the  marginal distribution of $Z$, an arbitrary observation from the longitudinal data in the latent space, with $n_i:=|\mathcal{I}_i|$ and  $n=\sum_{i\in \mathcal{I}} n_i$, is 
\begin{equation}
\label{eq:gaus_mixture_time_dep}
G_{Z}(z) = \frac{1}{n}\sum_{i\in\mathcal{I}}\sum_{j=1}^{n_i} \Phi\left(\frac{z-\mu_i(t_{i,j})}{\nu_i}\right).
\end{equation} 
So the marginal distribution of $Z$ is a Gaussian mixture over subjects and observation times.
The marginal variation across subjects, as in Section~\ref{sub:measures}, is captured exclusively through the distribution of the attributes $\{\alpha_i: i\in \mathcal{I}\}$.  All $\alpha_i$ are taken to be independent and identically distributed over subjects with $\alpha_i\sim N(0,V_{\alpha}^2)$ for all $i\in\mathcal{I}$, for a given fixed value of $V_{\alpha}>0$. 

From Section~\ref{sub:measures}, it is clear that ratio between the variance $V^2_{\alpha}$ of the $\{\alpha_i\}$ and the within-subject variance, i.e., $\nu^2_i$ for subject $i$, determines whether the longitudinal data exhibit asymptotic dependence or asymptotic independence. 
Hence $V_{\alpha}$ can be fixed to any chosen value, since the $\{\nu_i\}$ are 
estimated from the data, and so their values adapt proportionally to the choice of $V_{\alpha}$. 
Thus the data determine the form of longitudinal data extremal dependence.

\subsection{Transforming Margins between Observed and Latent Spaces}
\label{sec:transform}
The probability integral transform~\eqref{eq:PIT} links the observation scale of $X$ to and from the latent space of $Z$ defined in Sections~\ref{sub:modelling:subsec:popmodel} and \ref{sec:dep_modelling} respectively. For $F_{X_t}$ and $G_Z$ defined by expressions~\eqref{eq:GPD marginal_cdf} and \eqref{eq:gaus_mixture_time_dep}, respectively the variables $X_{i,j}$ and  $Z_{i,j}$, both at time $t_{i,j}$, are linked by 
\begin{equation}
    \label{eq:PIT}
    G_Z(Z_{i,j}) = F_{X_{t_{i,j}}}(X_{i,j}),\mbox{ so }
Z_{i,j} := T_t(X_{i,j})=G_Z^{-1}\{F_{X_{t_{i,j}}}(X_{i,j})\}
\end{equation}
for $T_t$ as in Section~\ref{sec:dep_modelling}.
For $X_{i,j}$ above the threshold on the original margins, the transform is 
\begin{equation}
\label{eq:PIT above}
Z_{i,j}= G_Z^{-1}\left\{1-\lambda_u(t_{i,j}) \left[1 + 
\xi (X_{i,j}-u)/\sigma_u(t_{i,j})\right]_+^{-\frac{1}{\xi}}\right\},
\end{equation}
whereas when these points are below the threshold,
$$Z_{i,j} = G_Z^{-1}\left\{\left[1-\lambda_u(t_{i,j})\right]\int_{-\infty}^{X_{i,j}} h_{t_{i,j}}(s)\diff s\right\}.$$
The threshold $u$ in the observation space becomes time-varying in the latent space, i.e.,  $u_Z(t)=G_Z^{-1}\left\{1-\lambda_u(t)\right\}$. 
As the density function $h_t$ is unknown and we do not want to model it, a censoring approach was proposed in Section~\ref{sec:Univariate}. For this range of $X_{i,j}$, the random variable $V_{i,j}:=\int_{-\infty}^{X_{i,j}} h_{t_{i,j}}(s)\diff s$
is uniform(0,1) distributed. So the auxiliary variable $V_{i,j}\sim \text{Uniform}(0,1)$ is introduced into the transformation when $X_{i,j}<u$, to give
$Z_{i,j} = G_Z^{-1}\left\{\left[1-\lambda_u(t_{i,j})\right] V_{i,j}\right\}.$

For making joint inferences across marginal and dependence structure parameters the likelihood functions in Section~\ref{sec: inference} require the Jacobian terms for these transformations. In each term the marginal density in the latent space is required, i.e.,
$$g_Z(z;\pmb{\theta},\pmb{\gamma},\pmb{\nu})=\frac{1}{n}\sum_{i\in\mathcal{I}}\sum_{j=1}^{n_i} 
\frac{1}{\nu_i}\phi\left(\frac{z-\mu_i(t_{i,j};,\pmb{\theta}_i,\pmb{\gamma})}{\nu_i}\right),$$
where $\pmb{\nu}:=\{\nu_i: i\in \mathcal{I}\}$ and $\pmb{\theta}:=\{\pmb{\theta}_i: i\in \mathcal{I}\}$.
For a realisation $x$ of $X$ (or $v$ of $V$) when the observation is above (or below) $u$, respectively, the associated realised value $z$ of $Z$ is obtained using the transformations above. 
For $\pmb{\sigma}$ and $\pmb{\beta}$ 
being parameters of the model for $\sigma_u(t)$ 
and $\lambda_{u}(t)$ respectively, the Jacobian terms at time $t$ for above $(J_+)$ and below $(J_{-})$ the threshold are
\begin{eqnarray}
\label{eq:Jacobian}
J_+(x;t, \xi, \pmb{\sigma}, \pmb{\beta}, \pmb{\theta},\pmb{\gamma},\pmb{\nu}) & =
& \frac{\lambda_u(t;\pmb{\beta})}{\sigma_u(t;\pmb{\sigma})
g_Z(z; \pmb{\theta},\pmb{\gamma}, \pmb{\nu})}
[1+\xi(x-u)/\sigma_u(t; \pmb{\sigma})]_+^{-\frac{1}{\xi}-1},\nonumber\\
J_- (v;t,\pmb{\beta}, \pmb{\theta}, \pmb{\gamma},
\pmb{\nu}) &=& [1-\lambda_u(t;\pmb{\beta})]/g_Z(z;\pmb{\theta},
\pmb{\gamma}, \pmb{\nu}).
\end{eqnarray}

\subsection{Predicting future extreme events in longitudinal data}
\label{sec:analytical results and simulation strategy (summary)}

In accounting for the longitudinal structure, predictions of extreme events regarding individual subjects are ascertainable, e.g., a new record by a particular subject $i\in \mathcal{I}$.
Such inferences incorporate each subject's mean function over time and temporal dependence,
with both aspects described by the Gaussian process model of Section~\ref{sec:dep_modelling}, which gives analytical solutions to such probabilities via closed form conditional distributions.
The supplementary material provides an example prediction, namely, the probability of a subject $i\in\mathcal{I}$ breaking the current record response $r$ in some future time period, with
the probability derived under an idealised scenario.

The evaluation of such probabilities under any realistic scenario is most simply conducted through Monte Carlo methods, simulating over different realisations of the longitudinal process for the fitted model. When subject-specific mean functions are non-constant, decaying eventually over time,  then in the longer-term the extreme events are more likely to be due to subjects not yet observed in $\mathcal{I}$. However, in the 
short-term these future extreme events are most likely to be obtained by current subjects in $\mathcal{I}$, followed by a transitional medium-term where extremes arise from a mixture of these populations of subjects. In the supplementary material we provide a simulation framework that integrates information across the three classes of future subjects: those subjects in $\mathcal{I}$, indexed by $\mathcal{I}^c$ with $\mathcal{I}^c\subseteq \mathcal{I}$, 
which are still producing at least one response above $u$ in the future time window; those subjects $\mathcal{I}^f$, which produced responses exclusively below the threshold within the observed time-frame and so $\{\mathcal{I}^f \cap \mathcal{I}\} = \emptyset$, but in the future produce a response above $u$; and those subjects $\mathcal{I}^n$ with no recordings at all within the observed time-frame but which in the future period produce at least one response above $u$.

\section{Inference}
\label{sec: inference}
The likelihood is constructed in two steps.
First, the parameters $(\xi, \pmb{\sigma},\pmb{\beta})$ and the vector of auxiliary variables for the marginal variables in the observed space are assumed known, so only the parameters affecting the latent space need to be estimated. Then the uncertainty in these marginal parameters and auxiliary variables is accounted for.
For deriving the likelihood in the latent space for a given subject $i$ with observations $(\pmb{Z}_i,\pmb{t}_i):=\{(Z_{i,j},t_{i,j}):j\in\mathcal{J}_i\}$,
we define
the correlation matrix between 
all of subject $i$'s observations by
 the correlation matrix $\Sigma_{\pmb{\kappa}}^{i}:= K_{\pmb{\kappa}}(\pmb{t}_i,\pmb{t}_i)$, i.e.,  
the $(j,k)^{th}$ entry $\Sigma_{\pmb{\kappa}}^{i,(j,k)} := K_{\pmb{\kappa}}(t_{i,j},t_{i,k})$ is the correlation between $Z_{i,j}$ and $Z_{i,k}$. As responses for a subject are from a multivariate Gaussian distribution and different subjects are independent, the likelihood in the latent space for responses $\pmb{z}:=\{\pmb{z}_i:i\in\mathcal{I}\}$ is 
\begin{equation}
    \label{eq:likelihood latent}
L_\ell\left(\pmb{z}; \pmb{t},\pmb{\theta},\pmb{\gamma},
\pmb{\nu},\pmb{\kappa}\right) \propto \prod_{i\in\mathcal{I}}
\nu_i^{-n_i}|\Sigma_{\pmb{\kappa}}^{i}|^{-1}\exp\left(-\frac{1}{2}\tilde{\pmb{z}}_i^T\Sigma_{\pmb{\kappa}}^{i}\;\tilde{\pmb{z}}_i\right)
\end{equation}
where $\pmb{t}:=\{\pmb{t}_i:i\in\mathcal{I}\}$ and $\tilde{\pmb{z}}_{i} := \left\{[z_{i,j}-\mu_i(t_{i,j};\pmb{\theta}_i, \pmb{\gamma})]/\nu_i:j\in\mathcal{J}_i\right\}$ for all $i\in\mathcal{I}$.
The full likelihood requires the Jacobian terms, from expression~\eqref{eq:Jacobian}, which control the transformations between the two spaces and account for parameters for the margins in the observational space being unknown. Let 
the sets of observations which are below and above the threshold be $\mathcal{L}_-:=\{(i,j):X_{i,j}\leq u:j\in\mathcal{J}_i, i\in \mathcal{I}\}$ and $\mathcal{L}_+:=\{(i,j):X_{i,j}>u:j\in\mathcal{J}_i, i\in \mathcal{I}\}$ respectively. 
The full likelihood of parameters $\pmb{\Theta}:= (\xi, \pmb{\sigma}, \pmb{\beta},\pmb{\theta},\pmb{\gamma}, 
\pmb{\nu},\pmb{\kappa})$ and auxiliary variables   
is 
\begin{align}
\label{eq:likelihood}
L(\pmb{x},\pmb{v}; & \pmb{t}, \pmb{\Theta})
\propto    
L_\ell\left(\pmb{z}; \pmb{t},\pmb{\theta},\pmb{\gamma}, 
\pmb{\nu},\pmb{\kappa}\right)\times
\nonumber\\
& \left(\prod_{(i,j)\in\mathcal{L}_-}J_-(v_{i,j};t_{i,j},\pmb{\beta},\pmb{\theta}, \pmb{\gamma},
\pmb{\nu})\right)
\left(\prod_{(i,j)\in\mathcal{L}_+}J_+(x_{i,j};t_{i,j},\xi,\pmb{\sigma},\pmb{\beta},\pmb{\theta},\pmb{\gamma},\pmb{\nu})\right),
\end{align}
where $\pmb{v}:=\{v_{i,j}:(i,j)\in\mathcal{L}_-\}$ and
$\pmb{z}$ is a function of $\pmb{x}$ and $\pmb{v}$, as identified in Section~\ref{sec:transform}.

 With two parameters per subject, limited data per subject, and many subjects, a asymptotic-based likelihood inference and its associated uncertainty evaluation is not supported. Avoiding such asymptotics via bootstrap sampling also has complications due to the auxiliary variables, and since subjects with limited data are likely to be omitted in replicate samples.
So, we adopt a Bayesian inference framework, which provides full uncertainty quantification of all parameters and auxiliary variables simultaneously. 

Let the parameters $\pmb{\Theta}$ have prior distribution $\pi_{\pmb{\Theta}}$, and let the prior $\pi_{V_{i,j}}(v)$ for all $(i,j)\in \mathcal{L}_-$ be uniform $(0,1)$ distributed and to be independent across these variables. Then, the full posterior distribution 
is $\pi\left(\pmb{\Theta}, \pmb{v}|\pmb{x},\pmb{t}\right) \propto \pi_{\pmb{\Theta}}\left(\pmb{\Theta}\right)
    L(\pmb{x},\pmb{v}; \pmb{t}, \pmb{\Theta})$.
In Section~\ref{sec:swimPrior} we present the prior 
$\pi_{\pmb{\Theta}}$ for our analysis of elite swimming data.

Inference and diagnostics were conducted using the Python package \textit{PyMC} \citep{salvatier2016probabilistic}, with the supplementary material containing more extensive computational details. To attain inference for future predictions  
the full prediction uncertainty is propagated through the inference. Given future simulated time-stamps $\pmb{t}^*_{i}$ of responses by a subject $i$, which are randomly generated by the process described in Section~\ref{sec:analytical results and simulation strategy (summary)}, the variables $Z_i(t) \sim \mathcal{GP}\left\{\mu(t; \pmb{\theta}_i,\pmb{\gamma}),K_{\pmb{\kappa}}
(\cdot,\cdot )\right\}$, are simulated jointly for $t$ over  $\pmb{t}^*_i$, 
for each 
random sample from the joint posterior $\pi\left(\pmb{\Theta}, \pmb{v}|\pmb{x},\pmb{t}\right)$. The sample is then transformed back to its original margins. 

For an observation below the threshold - which is by definition not extreme - the actual value on the original margins is unimportant for inference of extreme events. Only the time of occurrence and the knowledge that they are below the threshold are relevant.
However, for visualisation purposes it is useful to have some estimate of non-extreme values on the original scale, see Figure \ref{fig:posteriorpredictive}. In this case the empirical CDF is used, though it is acknowledged that this does not include the uncertainty in the distribution on the original margins.

\section{Application}
\label{sec: application}

\subsection{Data}

The data analysed constitutes mens’ 100m breaststroke results in FINA competitions in the period 2012-2019, obtained from the FINA website. 
Strategic decisions were made about which data to analyse. Only each swimmer's best time swam per competition was selected, i.e., one swim per competition; we chose to analyse negative swim-times, and then negate any estimated quantiles in order to provide results for actual swim-times; 
the threshold was selected as the 200th fastest personal best (PB) over the period 2001-18, giving the (negative) extreme threshold as $u=-61.125$ seconds; and we excluded data from all swimmers with $m \le 7$ swims.
The reasons for these choices are discussed in the supplementary material. The resultant data that we used for analysis contained 120 swimmers, with 1435 total responses.

\subsection{Modelling applied to swimming}
\label{sec: modelling swimming}
From findings in \cite{spearing2021ranking}, the conditional distribution of extreme swim-times $\Pr\{X<x|X>u\}$ for large $u$ can be treated as identically distributed over time, and so we take $\sigma_u(t)=:\sigma_u \in\mathbb{R}_+,\;\forall t$, i.e., $\pmb{\sigma}=\sigma_u$. The common temporal trend across the population of elite breaststroke swimmers can then be captured through the probability of exceeding the threshold $\lambda_u$, via a smooth 
monotonically increasing function for $\lambda_u$. A logit-linear functional form for $\lambda_u$ was found appropriate for the change in $\lambda_u$ over $t$. Specifically,
for a swim-time in year $t\in\{2012, \dots,2020\}$ and parameters 
$\pmb{\beta}:=(\beta_0,\beta_1)\in \mathbb{R}^2$, we take  
\begin{equation}
\label{eq:rate equation reparametrised}
\lambda_u(t;\pmb{{\beta}}) = \exp(\beta_0 + \beta_1 t)/[1 + \exp(\beta_0 + \beta_1 t)].
\end{equation}

In elite swimming, the subject-specific
trend captures a swimmer's \textit{career trajectory} - the tendency for athletes to enter elite sports as relatively inexperienced, improve until some individual \textit{peak} ability, and then decline before leaving the sport. Swimmers tend to improve rapidly towards their peak mean performance $\alpha_i$, at an age of $\tau_i$, as they mature physically, and then stop competing within a few years of reaching this peak. Here we allow the time at which peak mean performance is achieved to vary over swimmers to allow for their differences in maturity.
The lack of data in the decline of the career trajectory enables the parsimonious assumption of a symmetric career trajectory about the peak. From what can be identified from the data, after transformation to the latent space, a quadratic mean trend in age of swimmer, with curvature $\gamma<0$, seems a reasonable approximation to this mean performance progression. By including the covariate $b_i\in\mathbb{R}$, of swimmer $i$'s birth date, we have $t-b_i$, for $t>b_i$, as the age at which swimmer $i$ at time $t$. Thus, the mean function in latent space is     
$$
\mu_i(t;b_i,\pmb{\theta}_i,\gamma)= 
\alpha_i - \gamma(t-b_i-\tau_i)^2, \mbox{ for all }t\in \mathbb{R},$$
for all $i\in\mathcal{I}$, where 
$\pmb{\theta}_i =(\alpha_i,\tau_i) \in \mathbb{R}\times \mathbb{R}_+$, and here $\pmb{\gamma}=\gamma>0$. We have no swimmer-specific parameter for $\gamma$ given the limited number of swims per swimmer. There was no evidence
for variation over swimmers in their across-swim variability, so we took  $\nu_i =: \nu\in\mathbb{R}_+,\;\forall i \in\mathcal{I}$.  A different $\mu_i$ per swimmer seems sufficient to capture the across-swimmer effects.

\subsection{Prior specification}
\label{sec:swimPrior}
The supplementary material gives the DAG
for the model for this swimming application.   The priors are assumed to be mutually independent across all components of $\pmb{\Theta}$, i.e., 
\begin{align}  
\label{eqn:swimPrior}
\pi_{\pmb{\Theta}}\left(\pmb{\Theta}\right)=
\pi\left(\xi\right)\pi(\sigma_u)
\pi\left(\pmb{\beta}\right)
\left(\prod_{i\in\mathcal{I}}\pi\left(\alpha_i\right) \pi\left(\tau_i\right)\right) 
\pi\left(\gamma\right)
\pi\left(\nu\right)\pi\left(\pmb{\kappa}\right).
\end{align}
We now explain our choices of these marginal priors in the sequence shown in expression~\eqref{eqn:swimPrior}.

Discussion on priors for GPD parameters goes back to \cite{coles1996bayesian}. The shape parameter prior being ${\mbox{logit}\left(\xi+1\right)\sim \mathcal{N}(\text{logit}(0.8), 0.3)}$ restricts the domain of the shape parameter to $-1<\xi<0$. The constraint $\xi>-1$ avoids estimates of the GPD implying the best possible time has already been achieved, whilst $\xi<0$ imposes a finite limit on the fastest possible performance. Analysis of 2001-2019 elite swimmers' PB data 
found strong evidence of a common negative shape parameter for all swimming distances, strokes and gender categories \citep{spearing2021ranking}. For the GPD scale parameter prior we exploit knowledge from \cite{spearing2021ranking} 
that this parameter, estimated using PB data, was close to $1$: so $\sigma_u \sim \text{Gamma}(25,25)$ enforces positivity, has the required mean, and a standard deviation of $0.2$. For the threshold exceedance rate parameters $\pmb{\beta}$,
the priors $\beta_0\sim N(0,0.5)$, and $\beta_1 \sim \text{Gamma}(0.1,0.1)$ are imposed. The latter reflects the improvement of elite swimmers \citep{spearing2021ranking}, and when combined with the former gives exceedance rates in the range $(0.1,0.9)$.

Considering the priors for the latent space parameters,  
we take $\alpha_i {\sim} N(0,V^2_{\alpha})$, with $V_{\alpha}=6$.
The priors  $\tau_i\sim N\left(25,2.5^2\right)$ reflect that a swimmer's peak age is roughly 25 years, with a high probability of being in the interval $(17.5,32.5)$.  
The prior  $\gamma \sim \text{Gamma}(0.5,0.5)$ provides weak information with a preference for $\gamma$ to be close to 0, to ensure that a posterior with $\gamma>0$ is not a prior artefact. 
As it is anticipated there is greater variance between swimmers than within any swimmer's performances, so we take $\nu\sim\text{Gamma}(1,1)$ which has a smaller variance than $V^2_{\alpha}$. For the kernel parameters, taking $\kappa_0 \sim \text{Gamma}(0.5,0.5)$ and 
$\text{logit}(\kappa_1-0.5)/1.5) \sim N(\text{logit}(1),2)$ enforces $\kappa_0>0$ and  allows exploration over $\kappa_1\in(0.5,2)$.

\subsection{Results}
\label{results}
\subsubsection{Subject-specific Inference}

The within-subject features of the model
provide information about individual swimmers  as well as playing a key role in determining the dependence structure across  of the elite breaststroke swimmers. As identified in 
Section~\ref{sub:measures}, there are two features of the subject-specific behaviour which affect the extremal dependence of these data: the subject-specific variation in the attributes $\{\alpha_i: i \in \mathcal{I}\}$; and the within-subject dependence, given by the Gaussian process.

\begin{figure}[h!]
    \centering
     \begin{subfigure}[b]{0.325\textwidth}
         \centering
         \includegraphics[width=\textwidth]{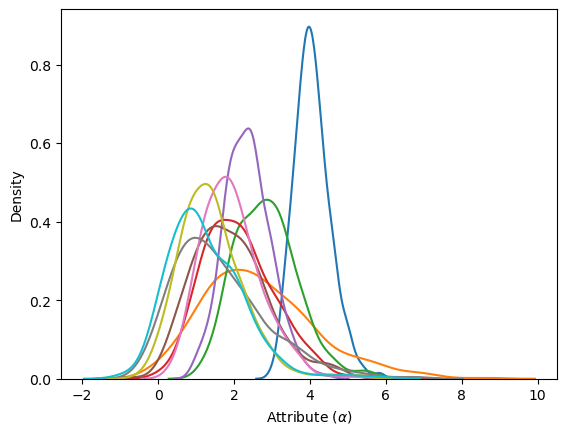}
         \label{fig:posterior inference:attribute}
     \end{subfigure}
      \begin{subfigure}[b]{0.325\textwidth}
         \centering
         \includegraphics[width=\textwidth]{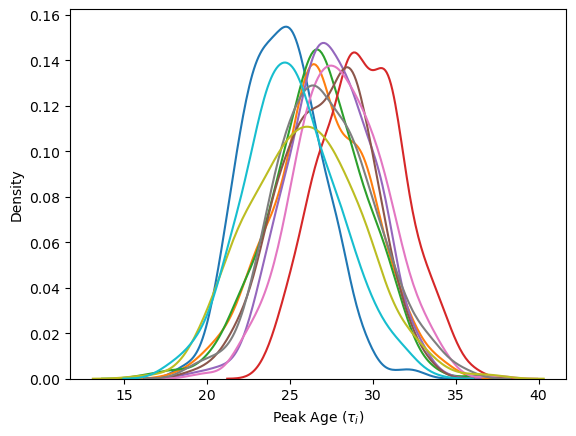}
         \label{fig:posterior inference: peak age}
     \end{subfigure}
     \hfill
       \begin{subfigure}[b]{0.325\textwidth}
         \centering
         \includegraphics[width=\textwidth]{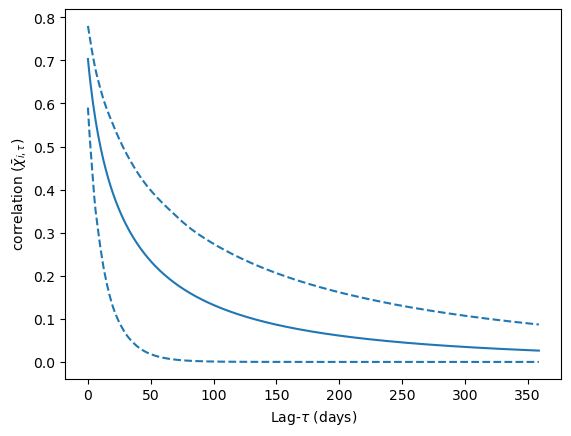}
        \label{fig:posterior inference:correlation}
     \end{subfigure}
     \hfill
    \caption{Subject-specific posterior inferences. For the top 10 swimmers, 
    the posteriors of these swimmers' attributes $\alpha_i$ (left) and peak ages $\tau_i$ (middle). The colours identify swimmers as defined in Figure~\ref{fig:individuals}~(left). The mean posterior and $95\%$ HPDI for the subject-specific asymptotic independence measure $\bar{\chi}_{i,\tau}$ against time lag $\tau$ in days (right). }
    \label{fig:posterior inference}
\end{figure}

The marginal posterior distributions of the parameters $\pmb{\theta}_i$ are shown in Figure~\ref{fig:posterior inference} for the top ten swimmers, as defined in Section \ref{sec:subjectPredictions}, a ranking that strongly correlates with the ten largest posterior mean $\alpha_i$ values. 
With the exception of the posterior for Adam Peaty's $\alpha_i$, there is considerable overlap between the other nine posteriors, with Peaty's having both a larger mean and 50\% of the variation of the others. The larger mean is not surprising as Peaty holds the 7 fastest times, and 11 of the top 20, for the competition-best data, together with all the top 20 times over all swims. The posteriors for
the $\tau_i$ for these swimmers are broadly more
self-consistent, with almost all posterior mass for the peak performance age in the range $(25,35)$ years, though both Peaty and Andrew Michael have lower peak ages, with Peaty almost certainly peaking before the age of 30 (he is 29 at the time of writing).

What is intriguing is that the posterior of $\alpha_i$ for Nicolo Martinenhi has upper quantiles which exceed the same quantiles for Peaty's $\alpha_i$, despite his  median being smaller than Peaty's. We explored three possible causes for this.
Firstly, it could be that Martinenhi produced  highly variable swim times, indicating that he is capable of better swims than Peaty; this is unlikely as only two of Peaty's swim-times are slower than Martinenhi's PB. Secondly, the posterior uncertainty of Martinenhi's $\alpha_i$   could be due to having less swims in the database relative to Peaty, but he has 14 better than the threshold, which is comparable to Peaty's 17. The  most likely, is that Martinenhi is relatively young - five years younger than Peaty - being aged 20 years in his most recent database entry. For younger swimmers it is difficult to disentangle between peak age and attribute, which is evidenced by 
Martinenhi having the largest posterior correlation, of $0.89$, between his $(\alpha_i,\tau_i)$ of the top ten swimmers, e.g., for Peaty this is $0.50$. Martinenhi's large uncertainty in peak age is contributing  to the uncertainty in his attribute; his peak is still to come - but we are uncertain in its level.

The posterior 95\% highest posterior density interval (HPDI) for the subject-specific quadratic trend curvature $\gamma$ is $(0.015, 0.029)$, showing that there is strong evidence of a rising and falling career trajectory, especially given the prior favours $\gamma$ being arbitrarily close to 0. 
The 95\% HPDI for the ratio of within-subject to across subject variation, i.e., $\nu/V_\alpha$, is $(0.17, 0.18)$, so the majority of the variation in the extremes of these longitudinal data is explained by swimmer identification. Furthermore, with Peaty having by far the largest $\alpha_i$, Section~\ref{sub:measures} indicates there will be asymptotic dependence, irrespective of the within-subject dependence $\rho(\tau)$ at lag $\tau$.
The posterior mean and pointwise 95\%  (HPDI) are shown in 
Figure~\ref{fig:posterior inference}~(right) for 
the measure of subject-specific asymptotic independence $\bar{\chi}_{i,\tau}=\rho(\tau)$, 
for lag $\tau\in [5,365]$ days. This inference indicates that at 50 days there is reasonable dependence per swimmer and even at 6 months lag there is non-negligible subject-conditional dependence.

\subsubsection{Subject-ignorant Marginal Inference}

The joint posterior inferences for the subject-ignorant marginal distribution parameters for the GPD and tail exceedance probabilities $(\sigma_u, \xi, \pmb{\beta})$ are derived from the full model joint posterior.
The posterior mean of $\xi$ and its 95\% HPDI are $-0.22~(-0.25,-0.20)$, they provide strong evidence for a negative shape parameter. For $\beta_1$ these values are $0.13~(0.09,0.16)$, showing that the rate of achieving extreme elite performances by swimmers indexed $\mathcal{I}$  is increasing over the time window, with the posterior mean and 95\% HPDI for $\lambda_u(t)$ being $0.34~(0.30,0.38)$ for 2012 and
$0.55~(0.51, 0.59)$ for 2019, a substantial difference in behaviour.

\begin{figure}[h!]
    \centering
     \begin{subfigure}[b]{0.48\textwidth}
         \centering
        \includegraphics[width=0.8\textwidth]{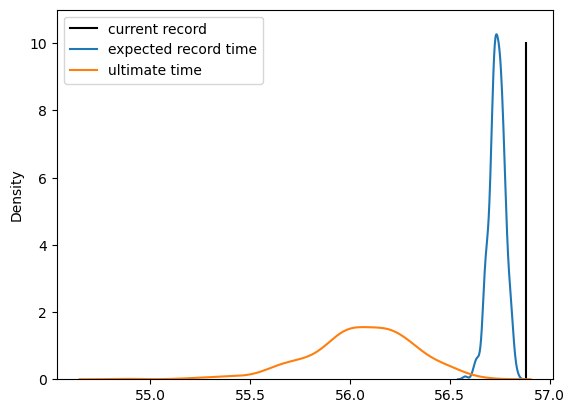}
     \end{subfigure}
      \begin{subfigure}[b]{0.48\textwidth}
         \centering
         \includegraphics[width=0.86\textwidth]{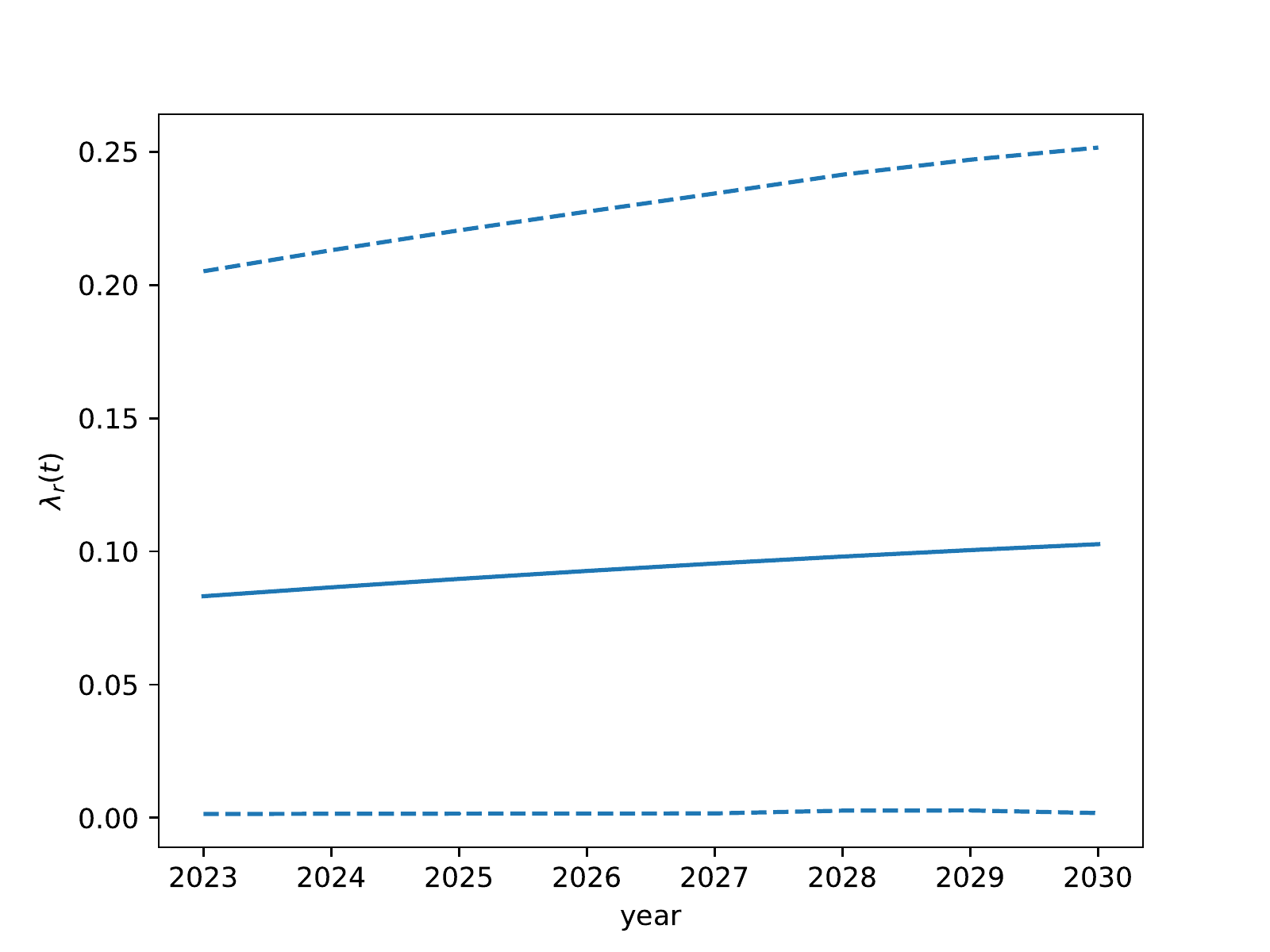}
     \end{subfigure}
    \caption{(Left: the posterior distributions for the expected next record swim-time (blue) and ultimate swim-time (orange)    
    for the mens' 100m breaststroke in seconds. Peaty's current record time  (black vertical line). Right: posterior mean (solid line) and 95\% HPDI (dashed lines) of the rate $\lambda_r(t)$ of swims by elite swimmers of beating Peaty's current record in year $t$.}
    \label{fig:ultimate time}
\end{figure}

As described in Section~\ref{sec:Theory}, when $\xi<0$ there is an estimated upper endpoint $x_H = u-\sigma_u/\xi$, which for swimming is the best performance humanly possible, given the current technology, in the event \citep{toussaint122005biomechanical, nevill2007there}.
Figure~\ref{fig:ultimate time} shows the posterior distribution of $x_H$, and the closeness of Peaty's current world record to this. The posterior places the endpoint closer to the current record than a similar analysis of PB data \citep{spearing2021ranking}, with that analysis pooling information across events.

The expected value of the next world record swim-time is obtained by exploiting the \textit{threshold-stability} property of a GPD \citep{coles2001introduction}. Since
the (negative) current world record $r=-56.88>u$, exceedances above $r$ follow a GPD, i.e., letting $X_{r_+}:=\{X:X>r\}$, then $X_{r_+}-r\sim
\mbox{GPD}(\sigma_r = \sigma_u + \xi(r-u),\xi)$ and the expected next world record time is $\mathbb{E}[X_{r_+}] = r + \sigma_r/(1-\xi)$. Figure~\ref{fig:ultimate time} (left) shows the posterior distribution of $\mathbb{E}[X_{r_+}]$. Although it has some overlap with the posterior of $x_H$, the posterior of $\mathbb{E}[X_{r_+}]$ is much nearer Peaty's current record than $x_H$. 
The simplicity of $\mathbb{E}[X_{r_+}]$
arises as both $\xi$ and $\sigma_u$ are constant over time and the expectation is not conditional on the current swimmers' performances, with the latter considered in 
Section~\ref{sec:subjectPredictions}. An indication about when this next record is likely to be achieved is given in Figure~\ref{fig:ultimate time} (right), where we present the posterior for the rate $\lambda_r(t)$ per future year $t$  of  swims by elite swimmers beating Peaty's record $r$. Here $\lambda_r(t)= s_t\lambda_u(t)[1+\xi(r-u)/\sigma_u]^{-1/\xi}$,
where $s_t$ is number of total swims per year by elite swimmers. The posterior mean and 95\% HPDI are shown for $\lambda_r(t)$ over the window $2023-30$, with $s_t=s_{2019}$ for $t>2019$.

\subsubsection{Model Diagnostics}
\label{sec:diagnostics}
\begin{figure}[h!]
    \centering
\includegraphics[scale=0.6]{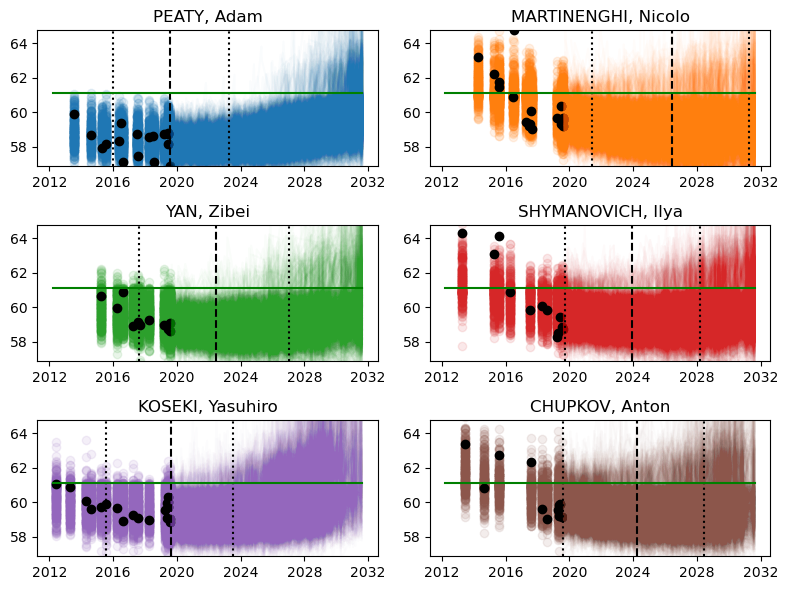}
    \caption{Within-subject diagnostics for six top swimmers: observed swim-dates and performance in seconds (black dots); posterior predictive distributions samples (coloured dots) for the dates of their swims in the past, and for future simulated swim dates. The threshold $u$ is  the horizontal line and the posterior mean and 95\% HPDIs for the peak age $\tau_i$ are  vertical lines.}
    \label{fig:posteriorpredictive}
\end{figure}

Diagnostics for the marginal GPD element of our model are well-established, so here novel diagnostics for the subject-specific characteristics of the data are presented. 
The diagnostics are shown on the observed scale, so observations can be compared with predictive distributions for the associated swim-dates.
Figure~\ref{fig:posteriorpredictive} shows 
the observations over time for six top swimmers, identified in Figure~\ref{fig:individuals}. All these swimmers have performances that are generally improving over time, and with some slower than the threshold. As such slow swims are treated as censored at the threshold,  modelling these precise values is not of great importance, with the prime focus concerning swim-times better than the threshold.

A sample size of 400 was generated from the posterior predictive sample for each past date of a swim for each of these swimmers.  
Figure~\ref{fig:posteriorpredictive} presents these samples under-laying the corresponding observations. In a well-fitting model, each observation should appear as a representative member of these samples.  The posterior predictive samples indicate that the model fits well, as most observations are reasonably central to their associated distribution for all swims better than the threshold, and 
even for the swims not as good as the threshold. They also capture the career trajectory evident in the data.
Maybe to be expected, Peaty's three best swim-times, each world records when achieved, are into the tails of their associated  predictive distributions. For weaker swims, 
Martinenghi and Shymanovich have performances which are unexpectedly slow relative to what our model would anticipate.

Figure~\ref{fig:posteriorpredictive}
also shows samples for these predictive distributions in the future, as the points from 2020-32, obtained under a stochastic model for the number and dates of future swims assuming that the swimmers continue to compete at current rates
(see the supplementary material for details). As most of these future samples improve or stay reasonably static over time, this illustrates that these swimmers  are early in their careers. In contrast, for Peaty there is a decay of performances from 2024.
On this figure are the posterior mean and 95\% HPDI for each swimmer's $\tau_i$, which cover the period where the predictive samples plateau.

\subsubsection{Subject-specific Predictions for Current 
Swimmers}
\label{sec:subjectPredictions}

Here we make predictive inference for future extreme events linked to specific swimmers,  thus illustrating the novelty of inferences that are possible using our longitudinal extreme value model. Section~\ref{sec:analytical results and simulation strategy (summary)} 
identified  three groups $(\mathcal{I}^c, \mathcal{I}^f, \mathcal{I}^n)$
of swimmers to consider when predicting future extreme events, and the supplementary material sets out the Monte Carlo strategies for the evaluation of the corresponding posterior distributions. To avoid the extra assumptions that are required to study groups $\mathcal{I}^f$ and  $\mathcal{I}^n$, only swimmers in $\mathcal{I}^c$ 
who have recordings in the most recent year of data are studied. 
From our model and posterior predictive inference,
standard extreme value properties, e.g., the distribution of the annual maxima, are simple to derive; however in sport, extreme events are mostly concerned with breaking records. Therefore, we focus on beating the current world record and setting PB times. Throughout, the future behaviour of swimmers is assumed consistent with the past data, so illness or sudden retirement are not accounted for, e.g.,  we ignore that Peaty has absences from the sport since 2021.

\begin{figure}
    \centering
    \begin{subfigure}[b]{0.25\textwidth}
         \centering
         \includegraphics[width=\textwidth]{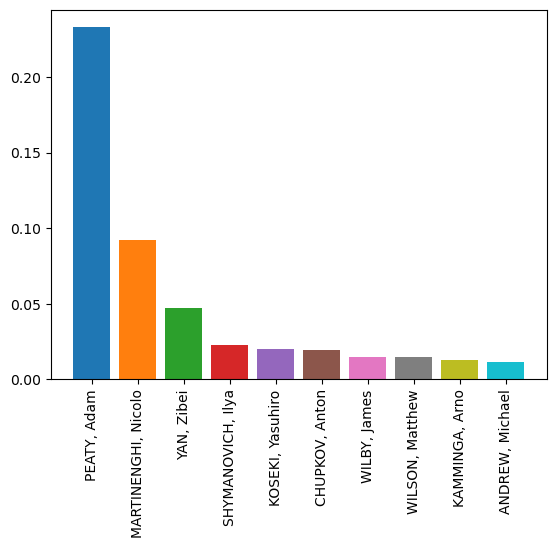}
         \label{fig:id}
     \end{subfigure}
     \hfill
     \begin{subfigure}[b]{0.35\textwidth}
         \centering
         \includegraphics[width=\textwidth]{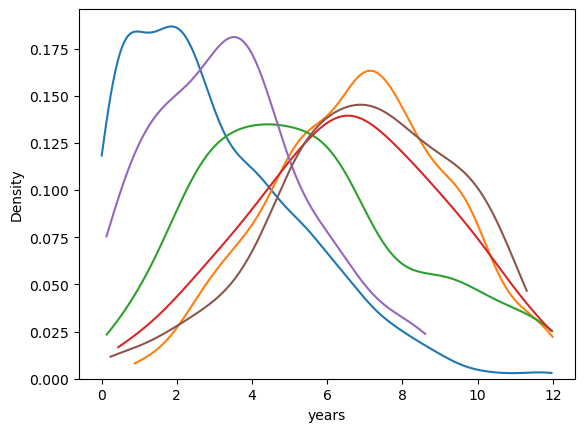}
         \label{fig:id_time}
     \end{subfigure}
      \begin{subfigure}[b]{0.35\textwidth}
         \centering
         \includegraphics[width=\textwidth]{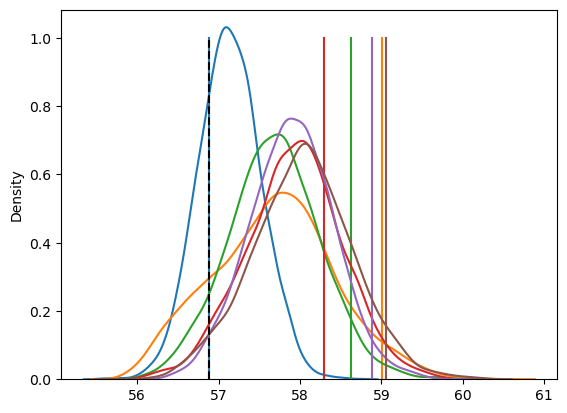}
         \label{fig:pb}
     \end{subfigure}
     \hfill
        \caption{Left: predictive probability that each swimmer will be the next swimmer in $\mathcal{I}^c$ to beat the current world record for the 10 most likely. Middle: the posterior distributions for each swimmer for the time at which they are the first the swimmers in $\mathcal{I}^c$ to beat the current record. Right: the posterior distributions of the expected PBs of all future times (vertical lines showing current PBs). 
        Swimmers are identified from the colours in left panel.}
        \label{fig:individuals}
\end{figure}

First consider the beating 
of the current world record. The joint posterior predictive distributions in Figure~\ref{fig:posteriorpredictive}, provide samples of future longitudinal data for the swimmers. 
There is a posterior predictive probability of $0.53$ that the world record is beaten by a swimmer in $\mathcal{I}^c$ in the next 12 years. 
The record will be found to be broken  with a larger probability in this window if we also account for the groups $\mathcal{I}^f$ or $\mathcal{I}^n$. 
Figure~\ref{fig:individuals} (left) splits this probability to show the posterior predictive probability for swimmer $i$ beating the record, for the 10 most likely swimmers in $\mathcal{I}^c$. This gives a novel 
\textit{ranking} method for swimmers within an event,
as it focuses on the future potential of swimmers (through accounting for their future career trajectory) more than their past achievement (the exclusive focus of typical ranking methods). 
Perhaps unsurprisingly, 
 Figure~\ref{fig:individuals} (left) shows that
Peaty is ranked the highest, i.e., the most likely to first beat his own world record of the swimmers in $\mathcal{I}^c$, with a predictive probability of $0.24$. 
Martinenghi is ranked second, as expected given Figure~\ref{fig:posterior inference}~(middle), with a predictive probability of $0.09$. 

To assess how soon these swimmers can first beat the current record,
Figure~\ref{fig:individuals} (middle) shows the predictive distribution of the year in which a swimmer will be the first of the current swimmers to beat the record. 
These posteriors are shown for the top six ranked swimmers in Figure~\ref{fig:individuals} (left). These results show that if Peaty does break his record, it is most likely to happen within the next four years, due to his age exceeding his peak age subsequently.
In contrast, Martinenghi is most likely to beat the current record in 4-10 years.
Figure \ref{fig:individuals} (right) shows the posterior distribution of the best time for each swimmer in the future window.
These distributions show that there is a reasonable chance of 
each swimmer beating their current PB. Peaty is less likely to do this than the other five swimmers shown, who all have a high posterior probability of beating their current PBs.
This finding is not surprising, as swimmers that are currently near their peak have a limited chance of beating their PBs whereas younger swimmers have the largest chance of setting new PBs as they are still improving.

\section{Discussion}
\label{sec: discussion}

This article proposes the first analysis for extreme values of data arising from a longitudinal structure comprising multiple subjects, each with a time series of responses. Although much new asymptotic theory remains to be developed, as the number of subjects and the lengths of their time series tend to infinity at potentially different rates, our focus has been in terms of putting down the framework for statistical modelling and associated inference. Furthermore, we have exhibited that this framework provides a basis for novel analysis of elite swimming data, and have illustrated the additional challenges that arise in practice, e.g., non-stationarity over subjects, subjects with very limited data, and the need to model subjects not in the data. 

This generic framework for longitudinal data analysis involving extreme values contains a set of modelling decisions which are application specific. Core examples are the choice of functional forms for the subject-specific mean function $\mu_i$ for all $i\in\mathcal{I}$, the threshold exceedance rate function $\lambda_u$, and the GPD scale parameter function $\sigma_u$. In our swimming application, 
fully parametric functional forms were established from prior application-specific knowledge. For the period of data we analysed, $\lambda_u$ was modelled to be monotonically increasing, reflecting knowledge that the quality of swimmers has been improving generally in this period. However, if data prior to 2010 were used, a monotonic form would be inappropriate due to the phasing out of performance-enhancing full-body swim-suits, see \cite{spearing2021ranking}. 

For swimmers with less than $m$ measurements, a decision must be made between including them all or 
discarding them from the analysis, at the cost of high computational inefficiency or bias respectively. Although we developed a pragmatic compromise, 
another possibility could cluster each subject with $m$ or less responses  with a subject with more than $m$ responses. Subjects in the same cluster would have a common $(\alpha_i,\tau_i)$ but different ages and performances. This approach benefits from using all data for inference, but it is still likely to be computational demanding given the complexity of cluster allocation when no simple rule is available.

An entirely novel aspect of our inference has been the subject-specific features such as the 
variation across subjects being modelled through attributes $\{\alpha_i:i\in\mathcal{I}\}$.
Although Gaussian marginals are leveraged on the grounds of the parsimony of conditional and unconditional Gaussian processes, this choice is rather unimportant to the outcomes of the inference. This is due to the weak common prior across attributes, resulting in a posterior which is driven by the data. 
The resulting posterior for a new subject's $\alpha_i$ is a Gaussian mixture model; where it is recognised that this reflects only subjects capable of achieving measurements above a high threshold, and is not applicable to the population as a whole. Despite this restriction to the extreme subjects, our analysis shows that the variation between attributes for swimmers is substantially larger than natural variation of extreme times for any selected swimmer. The analysis has disentangled the variations of the longitudinal data to better inform future inference for extremes and records, both unconditionally and conditionally, for the current elite swimmers. 
\section*{Acknowledgements}
Spearing gratefully acknowledges funding of the EPSRC funded STOR-i Centre for Doctoral Training (grant number EP/L015692/1), and ATASS Sports.

\bibliographystyle{apalike}
\bibliography{main}

\begin{thebibliography}{}

\bibitem[Andrieu and Roberts, 2009]{andrieu2009pseudo}
Andrieu, C. and Roberts, G.~O. (2009).
\newblock The pseudo-marginal approach for efficient {M}onte {C}arlo
  computations.
\newblock {\em The Annals of Statistics}, 37:697--725.

\bibitem[Coles, 2001]{coles2001introduction}
Coles, S.~G. (2001).
\newblock {\em An Introduction to Statistical Modeling of Extreme Values},
  volume 208.
\newblock Springer London.

\bibitem[Coles et~al., 1999]{coles1999dependence}
Coles, S.~G., Heffernan, J.~E., and Tawn, J.~A. (1999).
\newblock Dependence measures for extreme value analyses.
\newblock {\em Extremes}, 2:339--365.

\bibitem[Coles and Tawn, 1996]{coles1996bayesian}
Coles, S.~G. and Tawn, J.~A. (1996).
\newblock A {B}ayesian analysis of extreme rainfall data.
\newblock {\em Journal of the Royal Statistical Society Series C: Applied
  Statistics}, 45(4):463--478.

\bibitem[Davison and Smith, 1990]{davison1990models}
Davison, A.~C. and Smith, R.~L. (1990).
\newblock Models for exceedances over high thresholds (with discussion).
\newblock {\em Journal of the Royal Statistical Society: Series B},
  52(3):393--425.

\bibitem[de~Fondeville and Davison, 2022]{de2022functional}
de~Fondeville, R. and Davison, A.~C. (2022).
\newblock Functional peaks-over-threshold analysis.
\newblock {\em Journal of the Royal Statistical Society Series B},
  84(4):1392--1422.

\bibitem[Diggle et~al., 2002]{diggle2002analysis}
Diggle, P.~J., Heagerty, P., Liang, K.-Y., and Zeger, S. (2002).
\newblock {\em Analysis of {L}ongitudinal {D}ata}.
\newblock Oxford {U}niversity {P}ress.

\bibitem[Duane et~al., 1987]{duane1987hybrid}
Duane, S., Kennedy, A.~D., Pendleton, B.~J., and Roweth, D. (1987).
\newblock Hybrid {M}onte {C}arlo.
\newblock {\em Physics Letters B}, 195(2):216--222.

\bibitem[Dupuis et~al., 2023]{dupuis2023modeling}
Dupuis, D.~J., Engelke, S., and Trapin, L. (2023).
\newblock Modeling panels of extremes.
\newblock {\em The Annals of Applied Statistics}, 17(1):498--517.

\bibitem[Engelke and Hitz, 2020]{engelke2020graphical}
Engelke, S. and Hitz, A.~S. (2020).
\newblock Graphical models for extremes (with discussion).
\newblock {\em Journal of the Royal Statistical Society Series B},
  82(4):871--932.

\bibitem[Foug{\`e}res et~al., 2006]{fougeres2006pitting}
Foug{\`e}res, A.-L., Holm, S., and Rootz{\'e}n, H. (2006).
\newblock Pitting corrosion: Comparison of treatments with
  extreme-value--distributed responses.
\newblock {\em Technometrics}, 48(2):262--272.

\bibitem[Foug{\`e}res et~al., 2009]{fougeres2009models}
Foug{\`e}res, A.-L., Nolan, J.~P., and Rootz{\'e}n, H. (2009).
\newblock Models for dependent extremes using stable mixtures.
\newblock {\em Scandinavian Journal of Statistics}, 36(1):42--59.

\bibitem[Gelman and Rubin, 1992]{gelman1992inference}
Gelman, A. and Rubin, D.~B. (1992).
\newblock Inference from iterative simulation using multiple sequences.
\newblock {\em Statistical Science}, 7:457--472.

\bibitem[Gomes and Henriques-Rodrigues, 2019]{gomes2019swimming}
Gomes, D.~T. and Henriques-Rodrigues, L. (2019).
\newblock Swimming performance index based on extreme value theory.
\newblock {\em International Journal of Sports Science \& Coaching},
  14(1):51--62.

\bibitem[Heffernan and Tawn, 2004]{heffernan2004conditional}
Heffernan, J.~E. and Tawn, J.~A. (2004).
\newblock A conditional approach for multivariate extreme values (with
  discussion).
\newblock {\em Journal of the Royal Statistical Society: Series B},
  66(3):497--546.

\bibitem[Hoffman and Gelman, 2014]{hoffman2014no}
Hoffman, M.~D. and Gelman, A. (2014).
\newblock The no-u-turn sampler: adaptively setting path lengths in
  {H}amiltonian {M}onte {C}arlo.
\newblock {\em Journal of Machine Learning Research}, 15(1):1593--1623.

\bibitem[Huser and Wadsworth, 2019]{huser2019modeling}
Huser, R. and Wadsworth, J.~L. (2019).
\newblock Modeling spatial processes with unknown extremal dependence class.
\newblock {\em Journal of the American Statistical Association},
  114(525):434--444.

\bibitem[Huub and Trultens, 2005]{toussaint122005biomechanical}
Huub, T. and Trultens, M. (2005).
\newblock Biomechanical aspects of peak performance in human swimming.
\newblock {\em Animal Biology}, 55(1):17--40.

\bibitem[Laycock and Scarf, 1993]{laycock1993exceedances}
Laycock, P. and Scarf, P. (1993).
\newblock Exceedances, extremes, extrapolation and order statistics for pits,
  pitting and other localized corrosion phenomena.
\newblock {\em Corrosion Science}, 35(1-4):135--145.

\bibitem[Leadbetter, 1991]{leadbetter1991basis}
Leadbetter, M.~R. (1991).
\newblock On a basis for ‘peaks over threshold’ modeling.
\newblock {\em Statistics \& Probability Letters}, 12(4):357--362.

\bibitem[Leadbetter et~al., 2012]{leadbetter2012extremes}
Leadbetter, M.~R., Lindgren, G., and Rootz{\'e}n, H. (2012).
\newblock {\em {E}xtremes and {R}elated {P}roperties of {R}andom {S}equences
  and {P}rocesses}.
\newblock Springer Science \& Business Media.

\bibitem[Ledford and Tawn, 2003]{ledford2003diagnostics}
Ledford, A.~W. and Tawn, J.~A. (2003).
\newblock Diagnostics for dependence within time series extremes.
\newblock {\em Journal of the Royal Statistical Society: Series B},
  65(2):521--543.

\bibitem[Momoki and Yoshida, 2023]{momoki2023mixed}
Momoki, K. and Yoshida, T. (2023).
\newblock Mixed effects models for large sized clustered extremes.
\newblock {\em arXiv preprint arXiv:2305.05106}.

\bibitem[Nelsen, 2007]{nelsen2007introduction}
Nelsen, R.~B. (2007).
\newblock {\em An {I}ntroduction to {C}opulas}.
\newblock Springer Science \& Business Media.

\bibitem[Nevill et~al., 2007]{nevill2007there}
Nevill, A.~M., Whyte, G.~P., Holder, R.~L., and Peyrebrune, M. (2007).
\newblock Are there limits to swimming world records?
\newblock {\em International Journal of Sports Medicine}, 28(12):1012--1017.

\bibitem[Pickands, 1975]{pickands1975statistical}
Pickands, J. (1975).
\newblock Statistical inference using extreme order statistics.
\newblock {\em The Annals of Statistics}, 3(1):119.

\bibitem[Richards and Huser, 2022]{richards2022unifying}
Richards, J. and Huser, R. (2022).
\newblock A unifying partially-interpretable framework for neural network-based
  extreme quantile regression.
\newblock {\em arXiv preprint arXiv:2208.07581}.

\bibitem[Richards et~al., 2023]{richards2023joint}
Richards, J., Tawn, J.~A., and Brown, S. (2023).
\newblock Joint estimation of extreme spatially aggregated precipitation at
  different scales through mixture modelling.
\newblock {\em Spatial Statistics}, 53:100725.

\bibitem[Robinson and Tawn, 1995]{robinson1995statistics}
Robinson, M.~E. and Tawn, J.~A. (1995).
\newblock Statistics for exceptional athletics records.
\newblock {\em Journal of the Royal Statistical Society: Series C (Applied
  Statistics)}, 44(4):499--511.

\bibitem[Salvatier et~al., 2016]{salvatier2016probabilistic}
Salvatier, J., Wiecki, T.~V., and Fonnesbeck, C. (2016).
\newblock Probabilistic programming in {P}ython using {P}y{MC}3.
\newblock {\em PeerJ Computer Science}, 2:e55.

\bibitem[Scarrott and MacDonald, 2012]{scarrott2012review}
Scarrott, C. and MacDonald, A. (2012).
\newblock A review of extreme value threshold estimation and uncertainty
  quantification.
\newblock {\em REVSTAT--Statistical Journal}, 10(1):33--60.

\bibitem[Smith and Goodman, 2000]{smith2000bayesian}
Smith, R.~L. and Goodman, D. (2000).
\newblock {\em Bayesian Risk Analysis. Chapter 17 of Extremes and Integrated
  Risk Management, edited by P. Embrechts}.
\newblock Risk Books, London.

\bibitem[Southworth and Heffernan, 2012]{southworth2012extreme}
Southworth, H. and Heffernan, J.~E. (2012).
\newblock Extreme value modelling of laboratory safety data from clinical
  studies.
\newblock {\em Pharmaceutical Statistics}, 11(5):361--366.

\bibitem[Spearing et~al., 2021]{spearing2021ranking}
Spearing, H., Tawn, J.~A., Irons, D., Paulden, T., and Bennett, G. (2021).
\newblock Ranking, and other properties, of elite swimmers using extreme value
  theory.
\newblock {\em Journal of the Royal Statistical Society: Series A (Statistics
  in Society)}, 184(1):368--395.

\bibitem[Stephenson and Tawn, 2013]{stephenson2013determining}
Stephenson, A.~G. and Tawn, J.~A. (2013).
\newblock Determining the best track performances of all time using a
  conceptual population model for athletics records.
\newblock {\em Journal of Quantitative Analysis in Sports}, 9(1):67--76.

\bibitem[Strand and Boes, 1998]{strand1998modeling}
Strand, M. and Boes, D. (1998).
\newblock Modeling road racing times of competitive recreational runners using
  extreme value theory.
\newblock {\em The American Statistician}, 52(3):205--210.

\bibitem[Wadsworth and Tawn, 2022]{wadsworth2022higher}
Wadsworth, J.~L. and Tawn, J.~A. (2022).
\newblock Higher-dimensional spatial extremes via single-site conditioning.
\newblock {\em Spatial Statistics}, 51:100677.

\bibitem[Wadsworth et~al., 2010]{wadsworth2010accounting}
Wadsworth, J.~L., Tawn, J.~A., and Jonathan, P. (2010).
\newblock Accounting for choice of measurement scale in extreme value modeling.
\newblock {\em The Annals of Applied Statistics}, 4(3):1558--1578.

\bibitem[Winter and Tawn, 2017]{winter2017k}
Winter, H.~C. and Tawn, J.~A. (2017).
\newblock $k$th-order {M}arkov extremal models for assessing heatwave risks.
\newblock {\em Extremes}, 20:393--415.

\end{thebibliography}

\pagebreak
\begin{center}
\textbf{\large Supplementary material for \\``A framework for statistical modelling of the extremes of longitudinal data, applied to elite swimming"}
\end{center}
\setcounter{equation}{0}
\setcounter{figure}{0}
\setcounter{table}{0}
\setcounter{section}{0}
\setcounter{page}{1}
\makeatletter
\renewcommand{\theequation}{S.\arabic{equation}}
\renewcommand{\thefigure}{S:\arabic{figure}}
\renewcommand{\thesection}{S:\arabic{section}}
\renewcommand{\bibnumfmt}[1]{[S#1]}
\renewcommand{\citenumfont}[1]{S#1}

This document accompanies the article ``A framework for statistical modelling of the extremes of longitudinal data, applied to elite swimming'', and any references to sections, figures or tables refer to those in the main article, unless prefixed with ``S'', which then refers to items in this supplementary material.

Section \ref{sec: links with papers} discusses other works which concern longitudinal/panel data of extreme values. Section \ref{sec:relevant copula properties} provides more detail behind properties of copulas. Section \ref{sec: Additional dependence measures for longitudinal data} introduces measures of extremal dependence for longitudinal data. These measure are then explored in Section \ref{sec:sep:limit result}, which includes further investigations into the nature of the extremal dependence of the scenarios derived in Section~\ref{sub:measures}. 
Section~\ref{sec:analytical results} shows analytical results for probabilities of future extreme events in longitudinal data using the model of Section \ref{sec: modelling swimming}, under some simplifying assumptions. In reality, many applications will require to full flexibility of our novel model, as seen in Section~\ref{sec: application}, and in this case Monte Carlo simulation provides computational solutions. A strategy for this is set out in Section~\ref{sec:predictionsnewsubjects}. Section~\ref{sec:inference_issues} explores how the Bayesian inference is conducted in practice, including the choice of MCMC algorithm, and Section~\ref{sec: data pre-processing} pertains to the swimming application in Section~\ref{sec: application} of the main paper, detailing the data pre-processing steps specific to this data set.

\section{Links with other papers on panel data}
\label{sec: links with papers}
Even though the panel data analysis of \cite{dupuis2023modeling} and \cite{momoki2023mixed} suggests a considerable overlap with this set up, the focus of their modelling and inference is very different to ours, with their priority being marginal inference for different subjects whereas we infer the within-subject measurement dependence and a population-based marginal model. They consider a simplified setting where all $\mathcal{J}_i$ are equal to some common $\mathcal{J}$, and measurement dates are identical across all $i\in\mathcal{I}$. The data are split over $B$ blocks,
$\{\mathcal{J}^b: b=1, \ldots ,B\}$ forming a partition over $\mathcal{J}$. Then, the joint behaviour of the subject block maxima $\{\max_{j\in \mathcal{J}^b}X_{i,j}: \mbox{ for }i\in\mathcal{I}, b=1, \ldots ,B\}$ are studied assuming these are independent over blocks. The temporal dependence structure of the within-subject behaviour, which is the focus of our analysis, is not considered. Instead their focus is to pool/cluster subjects $\mathcal{I}$ into groups, where each group can be assumed to have a common marginal distribution.

Perhaps the closest approach to our modelling of longitudinal extremes is \cite{fougeres2009models}, who use a latent/random-effect positive stable mixture model to produce a multivariate extreme value distribution to model dependence in repeated observations of pit depth across different coupons. Since they assume all $X_{i,j}$ 
are conditionally independent and identically distributed given a random effect $R_i$ for each subject, the dependence across time per subject is exchangeable, i.e., all pairs $(X_{i,j}, X_{i,k})$, for $j\not= k \in\mathcal{J}_i$, have the same dependence structure. This limited form of temporal dependence per subject is likely be too simplistic for generic longitudinal data, where pairs $(X_{i,j}, X_{i,k})$ often have dependence weakening as the time between them, $|t_{i,j}-t_{i,k}|$, increases. 
The use of the positive stable distribution to capture the variation between subjects - through both the mean and variance of $X_{i,j}|R_i$, for all $j\in\mathcal{J}_i$ - leads to the largest $R_i$ values corresponding to the subjects with $X_{i,j}$ values that are much larger than for other subjects. This is highly restrictive both for the limitation on how the population is distributed, but also as it enforces a strong form of extremal dependence over time, termed \textit{asymptotic dependence}. Asymptotic dependence, defined in Section~\ref{sec:AsymptoticDependence}, constrains that if a subject gives the largest value in the population at some time point, then they are likely to do this at all time points.

\section{Relevant copula properties}
\label{sec:relevant copula properties}
\cite{fougeres2009models} use the copula of the multivariate extreme value distribution with logistic$(\alpha)$ dependence structure,  
which in the bivariate case has $(\chi,\bar{\chi})=(2-2^{\alpha},1)$ 
for $0\le \alpha<1$ and $(\chi,\bar{\chi})=(0,0)$ when $\alpha=1$.
In terms of extremal dependence, this copula model is restrictive as it cannot capture any positive dependence within the asymptotic independence case. It also has limitations for modelling longitudinal data: the copula is exchangeable, which is unrealistic for most time series data; and the conditional distributions for this copula are non-trivial to simulate from. The latter property complicates inference for future extreme events. Due to these features we instead consider the $d$-dimensional Gaussian copula 
\begin{equation}
    \label{eq:Gaussian_copula}
    C(\pmb{x}) = \int_{-\infty}^{\Phi^{-1}(x_1)}\cdots \int_{-\infty}^{\Phi^{-1}(x_d)} \phi_d(\pmb{s};\Sigma) \diff \pmb{s},
\end{equation}
for $\pmb{x}=(x_1, \ldots ,x_d)\in[0,1]^d$ and $\pmb{s} \in \mathbb{R}^d,$ with 
$\phi_d(\pmb{s};\Sigma)$ denoting the $d$-dimensional Gaussian density, with standardized margins and dependence structure determined by the $d\times d$ correlation matrix $\Sigma$. In the bivariate case this copula has the properties $(\chi,\bar{\chi})=(0,\rho)$ for correlation parameter $-1<\rho<1$, and $(\chi,\bar{\chi})=(1,1)$ for $\rho=1$ \citep{coles1999dependence}.
Furthermore, as the multivariate copula is determined by its bivariate marginals, which are all asymptotically independent (except for the pathological case when $\rho=1$), it is not necessary to consider asymptotic dependence at any higher order. 

\section{Additional dependence measures for longitudinal data}
\label{sec: Additional dependence measures for longitudinal data}
To study the extremal behaviour over subjects at each time point, consider $M_t:=\max_{i\in \mathcal{I}}X_{i}(t)$ for different $t$.
This leads to the lag $\tau$ dependence measure
\begin{equation*}
    \label{eq:AD measure - chi(M)}
    \chi^{(M)}_{\tau}:=\lim_{q\uparrow 1}\Pr(F^{(M)}(M_{\tau})>q \mid F^{(M)}(M_0)>q)
\end{equation*}
where $F^{(M)}(x):=\prod_{i\in \mathcal{I}} F_i(x)=\prod_{i\in \mathcal{I}} F(x; \alpha_i)$, and also its equivalent asymptotic independence measure
$\bar{\chi}^{(M)}_{\tau}$. An alternative is to consider dependence between values in the marginal tail for each time point. This corresponds to picking a random subject from the population $\mathcal{I}$ at each time point, giving the lag-$\tau$ dependence measure
\begin{equation*}
    \label{eq:AD measure - chi(R)}
\chi^{(R)}_{\tau}:=\lim_{q\uparrow 1}\Pr(F^{(R)}(X^{(R)}_{\tau})>q \mid 
F^{(R)}(X^{(R)}_{0})>q)
\end{equation*}
where $X^{(R)}_{\tau}$ is a random selection from 
$\{X_{i}(\tau):i\in\mathcal{I}\}$, so has marginal distribution function
$F^{(R)}(x):=\sum_{i=1}^n F(x;\alpha_i)/n$. Again the  
equivalent asymptotic independence measure is $\bar{\chi}_{\tau}^{(R)}$. When all subjects are identically distributed and have the same temporal dependence structure, then
each of these extreme dependence measures at lag-$\tau$ are identical to the  measure of asymptotic dependence 
(asymptotic independence)~\eqref{eq:extremal dependence} and \eqref{eq: coefficient asymptotic independence} respectively
for the associated identically distributed variables. 
Thus each measure has equal validity when assessing dependence for longitudinal data.

\section{Further limit results for studying extremal dependence of longitudinal data}
\label{sec:sep:limit result}
Building on the results from Section~\ref{sub:measures}, here we explore further the nature of extremal dependence in longitudinal data. To help better understand the asymptotic dependence case we consider a version of measure $\chi_{\tau}^{(M)}$ which allows both $n$ and the quantile to grow in combination.
Specifically,  consider the conditional probability
$\Pr(M_{n2}>x_n \mid M_{n1}>x_n)$,
where $x_n\rightarrow \infty$ and letting $\alpha_n=x_n-\delta$ for some constant $\delta$. The marginal probability is then
\begin{align*}
    \Pr(M_{n1}>x_n) = 1-\Pr(M_{n1}<x_n)=1-\left[\Phi(x_n)\right]^{n-1}\Phi(x_n-\mu_n).
\end{align*}
Now consider the joint probability
\begin{align*}
    \Pr(M_{n1}>x_n,M_{n2}>x_n) &= 1-\Pr(M_{n1}<x_n)-\Pr(M_{n1}<x_n)+\Pr(M_{n1}<x_n,M_{n2}<x_n)\nonumber\\
    &=1-2\left[\Phi(x_n)\right]^{n-1}\Phi(x_n-\alpha_n) + \left[\Phi(x_n)\right]^{n-1}\Phi_2(x_n-\alpha_n, x_n-\alpha_n; \rho).
\end{align*}
Then, in case (i), consider setting $\alpha_n$ as above with $x_n$, this gives the limit
\begin{align*}
   & \Pr(M_{n2}>x_n\mid M_{n1}>x_n) 
    \rightarrow\frac{1-2\Phi(\delta) + \Phi_2(\delta, \delta; \rho)}{1-\Phi(\delta)}.
\end{align*}
The above limit is non-zero for all finite $\delta$ and when $\rho=0$ this limit is  $1-\Phi(\delta)$, which is positive for all $\delta<\infty$. So, when $\rho=0$, despite the independence of within-subject observations, the longitudinal structure induces asymptotic dependence.
 This is different from the findings for $\rho=0$ in limit~\eqref{eq:deplimitNormal}, showing the limits that give identical findings about the form of extremal dependence for identically distributed variables can give contrary results for longitudinal data. For case (ii) we have that
$\Pr(M_{n2}>x_n\mid M_{n1}>x_n)\rightarrow 0$, i.e. asymptotic independence.

Underlying all these limiting results is the fact that subject $n$ will be the componentwise maximum with probability 1 in case (i) and 0 in case (ii) for how $\alpha_n$ grows. This is shown through the following limit, which for case (i) explores the probability that the same subject gives a large measurement value at each time point, i.e.,
\begin{align}
     &\Pr(X_{n1}=M_{n1}, X_{n2}=M_{n2}\mid  M_{n1}>\alpha_n) \nonumber\\
     &=\Pr\{\max(X_{11},\dots,X_{(n-1)1})<X_{n1},X_{n1}>\alpha_n,\max(X_{12},\dots,X_{(n-1)2})<X_{n2}\mid  M_{n1}> \alpha_n\}\nonumber\\
     &=\int_{y-\infty}^\infty \int_{x=0}^\infty\Pr\{
     \max(X_{11},\dots,X_{(n-1)1})<\alpha_n+x,
     \max(X_{12},\dots,X_{(n-1)2})<\alpha_n+y \nonumber \\ 
     & \qquad \qquad \qquad \qquad \qquad  
     \mid   X_{n1}=x,X_{n2}=y\}\phi_2\left(x,y;\rho\right)\diff x\diff y
     /\Pr(M_{n1}> \alpha_n)\nonumber\\
&= \int_{y=-\infty}^\infty \int_{x=0}^\infty [\Phi(\alpha_n+x)
     \Phi(\alpha_n+y)]^{(n-1)}
     \phi_2\left(x,y;\rho\right)\diff x\diff y/\Pr(M_{n1}> \alpha_n)\nonumber\\
&\rightarrow 2\int_{y=-\infty}^\infty \int_{x=0}^\infty 
     \phi_2\left(x,y;\rho\right)\diff x\diff y= 1
\end{align}
as $n\rightarrow \infty$, as the powered terms tend to $1$, as in Section~\ref{sub:measures}, and that limit with $x=0$ explains the denominator tending to $1/2$, and the double integral is $1/2$ due to symmetry of the standard bivariate normal density about $x=0$.
Similarly, for case (ii) this limit is 0.

\section{Evaluation of probabilities of future extreme events for longitudinal data}
\label{sec:analytical results}

A benefit of accounting for the longitudinal structure is that now inference and predictions of extreme events regarding individual subjects is ascertainable, e.g., the probability that a new record is achieved  by a particular subject $i\in \mathcal{I}$. To make such inferences, each subject's mean function over time is incorporated, as well as the temporal dependence around this. Both of these aspects are described by the Gaussian process model of Section~\ref{sec:dep_modelling}, which gives analytical solutions to probabilities of future events through its closed form conditional distributions. Let $\mathcal{J}^{T_p}_i$ be the set of future measurements
for subject $i$, with the future schedule of measurement points defined as $\{t_{i,j}: j \in\mathcal{J}^{T_p}_i, i\in \mathcal{I}^{T_p}\}$, where all such $t_{i,j}>t_{\max}$ for a current time $t_{\max}$.
 
In practice the evaluation of the probabilities of such complex events are most simply conducted through Monte Carlo methods, simulating over different realisations of the longitudinal process for the fitted model, with evaluation achieved empirically over a large sample of replicates.
We present results and various assumptions of this type in Section~\ref{sec:predictionsnewsubjects}, but here we derive the analytical expression for one such event under an idealised set-up to illustrate the complexity even in this simplified scenario.

Consider the event $A^{T_p}_i(r)$, corresponding to the subject $i \in\mathcal{I}$ breaking the record for the maximum measurement in some future time period identified by ${T_p}$ and holding that record at the end of period, given that the current maximum measurement is $r$.  Consider the case where
(i) all parameters of the model in the latent space are known;
(ii) no subjects outside $\mathcal{I}$ produce measurements in time  period ${T_p}$; (iii) the observed subjects
have a constant mean function over time, i.e., 
$\mu_i(t)=\alpha_i$ in expression~\eqref{eqn:measurefunction};  and (iv) that there is subject-conditional independence for each subject. A benefit of 
assumption~(iv) is that it removes the need to consider the history of each subject's measurements including which subject holds the current record.
 
 To derive $P(A^{T_p}_i(r))$ it is most easy to work in the latent space, recognising that the current record transforms to the value $r_Z:= G_Z^{-1}\left[F_Z\left(r, t_r\right)\right]$ in  the latent space. First define $M^{T_p}_i:=\max\left(\{Z_{i,j}: j\in \mathcal{J}^{T_p}_i\}\right)$, the maximum measurement for subject $i$ in the future time period, then this distribution has the survivor function of $P(M^{T_p}_i>z)=1-[\Phi(z;\alpha_i,\nu_i)]^{\mid \mathcal{J}^{T_p}_i\mid}$, given assumptions~(iii) and (iv). Also let 
 $M^{T_p}_{-i} = \max\left(\{Z_{kj}: j\in\mathcal{J}^{T_p}_{k}, k\in\mathcal{I}\setminus \{i\}\}\right)$ be the maximum of all other subjects' measurements in this future period.
Then $P(A^{T_p}_i(r))$ is given as
\begin{align}
P(A^{T_p}_i(r)) =& P\{M^{T_p}_i > \max[r_Z,M^{T_p} _{-i}]\} 
=P\{M^{T_p}_i > r_Z> M^{T_p}_{-i}]\}+ P\{M^{T_p}_i >  M^{T_p}_{-i}> r_Z\}
\nonumber\\
=&
P(M^{T_p}_i>r_Z)
\left(\prod_{k\in\mathcal{I}\setminus\{i\}}\Phi(r_Z;\mu_k,\nu_k)^{|\mathcal{J}_k|} \right)+\int_{r_Z}^\infty 
P(M^{T_p}_i>z)f_{M^{T_p}_{-i}}(z)\diff z,\label{eq:analytica result}
\end{align}
where 
$$f_{M^{T_p}_{-i}}(z) = \left(\prod_{h\in\mathcal{I}\setminus\{i\}}\Phi(z;\mu_h,\nu_h)^{|\mathcal{J}_h|}\right)
\sum_{k\in\mathcal{I}\setminus\{i\}} |\mathcal{J}_k|
\frac{\phi(z;\mu_k, \nu_k)}{\Phi(z;\mu_k, \nu_k)}.$$

\section{Adapting predictions for new subjects}
\label{sec:predictionsnewsubjects}
For making inferences about  the future behaviour of extreme values for longitudinal data there are a number of substantial challenges linked to the subject-specific characteristics of the data structure. Analytical results such as result~\eqref{eq:analytica result} are available in simple cases, but with the mean functions inducing non-identically-distributed variables, it must be recognised that, in the longer-term, the extreme events are more likely to be due to subjects not yet observed in $\mathcal{I}$. In the short-term however,
these future extreme events are most likely to be obtained by the current subjects in $\mathcal{I}$, followed by a transitional medium-term in which extremes arise from a mixture of these populations of subjects. Here we develop the outline of a framework for such inferences, setting out some possible  choices that need to be made in relation to the currently unobserved subjects. The model parameters here are treated as known, and Section~\ref{sec: inference} presents how to account for that additional uncertainty. 

For the observed data there are $n$ subjects, indexed $\mathcal{I}$, each with at least one measurement above the threshold $u$. Going forward beyond the observed time-frame,
there are then three types of subject to consider: (i) those subjects in $\mathcal{I}$, indexed by $\mathcal{I}^c$ with $\mathcal{I}^c\subseteq \mathcal{I}$, 
which are still producing at least one measurement above $u$ in the future time window; (ii) those subjects $\mathcal{I}^f$, which produced measurements exclusively below the threshold within the observed time-frame and so $\{\mathcal{I}^f \cap \mathcal{I}\} = \emptyset$, but in the future produce a measurement above $u$; and (iii) those subjects $\mathcal{I}^n$ with no recordings at all within the observed time-frame but which in the future period produce at least one  measurement above $u$. To help remember the terminology the superscripts here denote $c$ for {\it current} subjects with a future threshold exceedance, $f$ for subjects in the population which are active in the observed time-frame and which record their {\it first} exceedance of $u$ in the future time period, and $n$ for an entirely {\it new} subject which records an exceedance of $u$ in the future time period.

For each subject in each of the groups $\mathcal{I}^c, 
\mathcal{I}^f$ and $\mathcal{I}^n$ measurement series are simulated over a time window of $(t_{\max}, t_{\max}+T)$ where $t_{\max}$ is the maximum time in the observed database and $T$ is the length of the future period of interest. 
As membership of these three groups depends on a subject achieving a measurement larger than $u$ in the future time-period, the number in each group is random. In practice it is easiest to first generate a time series for each individual that could be in the three groups and then a random number of these will meet the criteria to be in the respective groups.
For groups $\mathcal{I}^c$ and $\mathcal{I}^f$ the maximum number of potential subjects there could be is known from the observed numbers in the database, but in practice, computational time is saved by omitting previously measured subjects which have no measurements in the latter part of the observation window. That is, $t_{\max}-t_{i,n_i}$ being sufficiently large suggests that subject $i$ has stopped generating measurements that have potential to be extreme.
In contrast, for  $\mathcal{I}^n$ assumptions must be made about the arrival rate of new potential subjects. We propose that the rate of first measurements per subject in the database per unit time-period, denoted by  $r_{data}$, is used to estimate this rate. Then, the number of potential new subjects for the future is generated by a Poisson$(Tr_{data})$ random variable.

For each of the potential subjects in the three groups, the number and the times of the future measurements in $(t_{\max}, t_{\max}+T)$ and simulated realisations of the associated measurement 
$X_{i,j}$ are generated according to the models in Sections~\ref{sub:modelling:subsec:popmodel} and \ref{sec:dep_modelling} for these times-points, conditional on any information already present about these subjects. The subject is then identified as being from a group if their  maximum measurement exceeds $u$.
These steps are discussed below, identifying the features that change across the three groups.

For each potential subject in any of the three groups, measurement time points are generated independently over subjects from a homogeneous Poisson process with rate $\omega_i$ per unit time for subject $i$. That is, a subject $i$ has $N_i$ future observations, with  $N_i\sim 
\text{Poisson}(T\omega_i)$, with  these measurement time-points uniformly distributed on $(t_{\max}, t_{\max}+T)$. 
The times for the future measurements 
for subject $i$ are denoted by $\pmb{t}^*_{i,j}:=\{t^*_{i,j}: t_{\max}<t^*_{i,j}< t_{\max}+T, i=1, \ldots n^*_i\}$ where $n^*_i$ is the realisation of $N_i$.
For a potential subject $i\in \{\mathcal{I}^c \cup \mathcal{I}^f\}$, an estimate of $\omega_i$ is based on the empirical rate of measurements up to $t_{\max}$ for the subject in the database. 
For each potential subject $i\in \mathcal{I}^n$, $\omega_i$ is estimated from the observed population of subjects $\mathcal{I}$. Specifically, a subject $j$ is randomly drawn from 
$\mathcal{I}$ with associated rate $\omega_j$, and then we take $\omega_i\sim \mbox{log}N((\log(\omega_j)-\psi^2/2, \psi^2)$. This choice ensures that the expected value of $\omega_i$ is an existing rate $\omega_j$, and where the choice of $\psi$ can be selected based on how representative 
the subjects in $\mathcal{I}$ are believed to be relative to 
the entire population. So, $\psi$ can be taken larger if an under-representation of $\mathcal{I}$ is anticipated.

Next, the measurement values for each potential subject are simulated in the latent space, given the simulated future measurement times. 
For each potential subject $i$, measurements are simulated from the Gaussian process $Z_i(t) \sim \mathcal{GP}\left\{\mu(t; \pmb{\theta}_i,\pmb{\gamma}),K_{\pmb{\kappa}}
(\cdot,\cdot )\right\}$, at time-points
$\pmb{t}^*_{i}=(t^*_{i,1},\ldots  ,t^*_{i,n^{*}_{i}})$. These simulated processes are generated conditionally on the previous data when appropriate for the group, see below.
For future realisations in the upper tail of the latent variable space we can transform back to the observed space using transformation~\eqref{eq:PIT}. Only those potential subjects with their maximum measurement in the time interval
$(t_{\max}, t_{\max}+T)$ exceeding $u$ are included as a subject in their respective group. For deriving future scenarios we are only interested in those simulated measurements above $u$ in original space.

We have different existing knowledge at time $t_{\max}$
for each subject depending on which of the three groups they are from, in the form of past measurement values, covariates, and information about $\pmb{\theta}_i$. For a potential subject $i\in \mathcal{I}^c$, the posterior distribution of $\pmb{\theta}_i$ and the subject's covariates that determine how $\mu_i$ varies with $t$ are available. That potential subject's Gaussian process is the simulated given the past values of $(Z_i(t_{i,1}),\ldots  ,Z_i(t_{i,n_{i}}))$. Although some of these past $Z_i(t)$ values are non-extreme in the original space, i.e., the associated $X_{i,j}<u$, our inference methods of Section~\ref{sec: inference} provide estimates for all of these values from which to condition on for each of the generated posterior samples for the model parameters.

Now consider a potential subject $i\in \mathcal{I}^f$. Although past observational data are available for them, as of time $t_\text{max}$ these data are not included in inference, and so no estimates of subject-specific parameters $\pmb{\theta}_i$ are available. Likewise for any potential subject in group $\mathcal{I}^n$. For both cases $\pmb{\theta}_i$ can be drawn from the joint posterior from a randomly selected subject in $\mathcal{I}$. 
In both cases the Gaussian process is simulated forward from $t_{\max}$ independent of any past measurement data information, so for potential subjects in $\mathcal{I}^f$ the past data is ignored. To be able to use the Gaussian process, the relevant covariates for the potential subject are required. For
a potential subject $i\in \mathcal{I}^f$ their actual covariates are used, whereas for $i\in \mathcal{I}^n$ the covariates at drawn randomly from a subject in the database (not just from subjects in $\mathcal{I}$).

\section{Numerical Issues affecting Inference}
\label{sec:inference_issues}

Here we identify a computational issue that influences our choice of MCMC strategy. Specifically, in order to transform the data from the observed space into the latent space, the inverse of the Gaussian mixture distribution~\eqref{eq:gaus_mixture_time_dep} is required, but that has no analytical solution. 
Numerical solution of this inverse is required
for each likelihood evaluation, and for each data point for each subject. Exact numerical solution on this scale is computationally infeasible. Instead, for likelihood evaluation we use a grid search algorithm, searching over a finite regular grid $\mathcal{Z}_G$ in the latent space, for each data point $x_{i,j}$,  such that 
\begin{equation}
    \label{eq:grid search}
    z_{i,j}=
    \left\{\begin{matrix*}
    \argmin_{z \in \mathcal{Z}_G} \left(|G_Z(z) - F_X\left(x_{i,j},t_{i,j}\right)|\right), & x_{i,j}>u,\\
        \argmin_{z \in \mathcal{Z}_G} \left(|G_Z(z) - [1-\lambda_u(t_{i,j})]v_{i,j}|\right), & x_{i,j}\leq u,
    \end{matrix*}\right.
\end{equation} 
where both $F_X$ and $G_Z$ depend on the parameter values of each likelihood evaluation. This grid search approach slows down inference significantly since it requires a factor of $|\mathcal{Z}_G|$ more evaluations relative to there being an exact solution to equation~\eqref{eq:PIT}.
Moreover, the discrete nature of the grid search, with no gradient information, rules out our use of a range of popular Bayesian inference algorithms, e.g., Hamiltonian Monte Carlo \citep{duane1987hybrid} and the No U-Turn Sampler \citep{hoffman2014no}. Section~\ref{sec: discussion} discusses this point further.

Given these constraints and the slow likelihood evaluation, a Metropolis-Hastings (MH) algorithm is implemented that utilises the Python package \textit{PyMC} \citep{salvatier2016probabilistic}, which enables efficient inference through automated optimisation of the algorithms' tuning parameters. In the case of MH, this provides well-tuned proposal distributions for optimal exploration of the joint posterior distribution. For a further speed-up, we sample a large number (in our case 40) of shorter MCMC chains in parallel using high-performance computing, which then undergo standard diagnostics for checking of convergence \citep{gelman1992inference}.

By randomly drawing all realisations $\pmb{v}$ from its prior distribution at each step of the MCMC algorithm, and then considering only the marginal distribution for $\pmb{\Theta}$, in essence pseudo-marginal MCMC \citep{andrieu2009pseudo} is performed, and the posterior $\pi\left(\pmb{\Theta}|\pmb{x},\pmb{t}\right)$ is recovered.

\begin{figure}[h]
    \centering
    \input{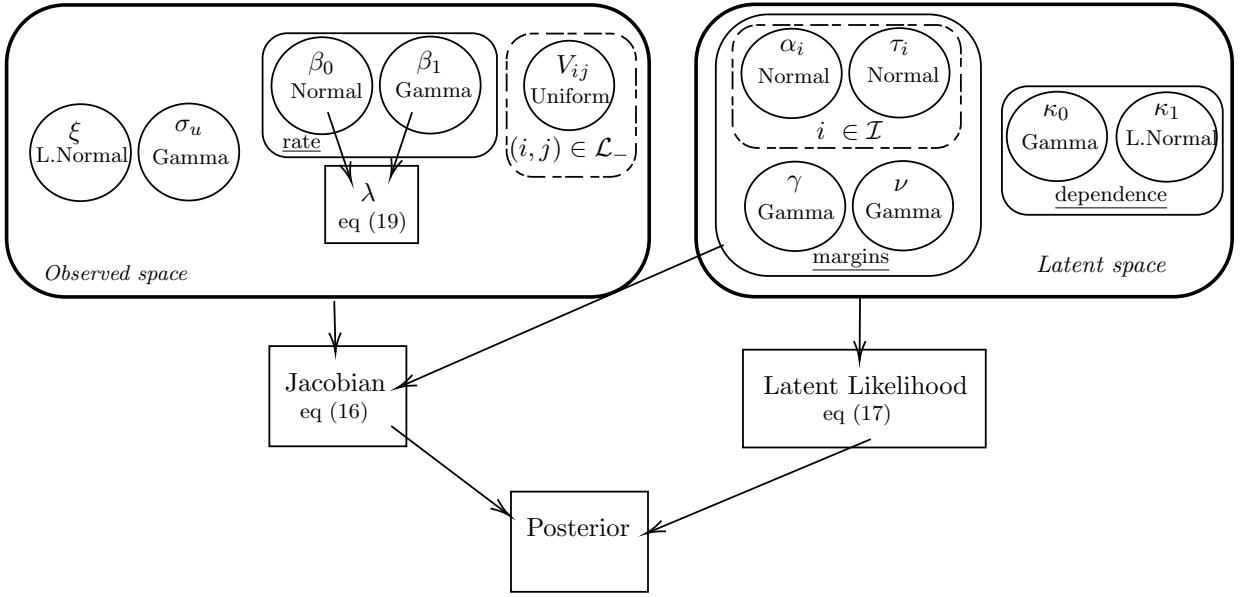}

    \caption{DAG illustrating the model flow with associated priors. The \textit{observed space} (left) shows the parameters for the extreme margins: $(\xi, \sigma_u)$, the GPD parameters; $(\beta_0,\beta_1)$, of the rate function $\lambda_u$ for exceeding the threshold $u$; and the auxiliary variables $V_{i,j}:(i,j) \in\mathcal{L}_-$ corresponding to the censored observations below $u$. The parameters of the \textit{latent space} (right), $(\pmb{\theta}:=\{\pmb{\theta}_i :=(\alpha_i,\tau_i):i\in\mathcal{I}\}, \gamma, \nu)$, determining the marginal distribution of the Gaussian mixture, and the kernel parameters $\pmb{\kappa}:=(\kappa_0, \kappa_1)$ which dictate the dependence structure.
    Both the observed space and latent space parameters determine the Jacobian \eqref{eq:Jacobian}, whilst the only the latent space parameters determine the latent-likelihood \eqref{eq:likelihood latent}. The posterior distribution then contains the Jacobian, the latent-likelihood, and the prior distributions 
    \eqref{eqn:swimPrior}.}
    \label{fig:Dag}
\end{figure}

The DAG in Figure~\ref{fig:Dag} illustrates the full model specification for the swimming application of Section~\ref{sec: application}, and
in particular, the formulation of the posterior.

\tikzset{every picture/.style={line width=0.75pt}} 

\section{Further details of the data pre-processing}
\label{sec: data pre-processing}
A few strategic decisions were made about which data to analyse. Firstly, only data of each swimmer's best time swam per competition was selected, i.e., one swim per competition. This removes much of the tactical element, e.g., weaker swimmers may need to swim to full capacity during the heats of competitions, whereas a top swimmer can typically afford to save their best performances for the finals. Using exclusively these \textit{competition maxima} helps to ensure that each observation is a good approximation of the swimmer's best ability at that time. It also has the benefit of avoiding the need to capture performance strategy or to deal with issues of dependence at very short time lags.

Secondly, in extreme value analysis, the scale on which the data analysis is performed can impact the results \citep{wadsworth2010accounting}. Following the discussion in \cite{spearing2021ranking} minimum swim-times are modelled, but modelling the maximum swim-speed \citep{gomes2019swimming}, i.e., the reciprocal of the times swam, is also an option. For analysing minimum swim-times, results exist for the behaviour of the lower tails of a distribution, however they are rarely applied \citep{robinson1995statistics} and give identical results to our strategy. We therefore analyse negative swim-times, and then negate any estimated quantiles in order to provide results for actual swim-times that make use of the more commonly-used methodological frameworks for upper tails. 

Next, the threshold must be selected. EVT gives that the generalised Pareto distribution (GPD) is  the only non-degenerate limit distribution for scale-normalised difference of $X$ from the threshold $v$ as $v$ tends to the lower endpoint of the distribution of $X$ \citep{pickands1975statistical}. In practice it is common to assume that the GPD is a sufficiently good approximation to the data relative to the threshold \citep{davison1990models}. The analysis of \cite{spearing2021ranking} identified that the personal-best swim-times better than the 200th top personal-best time, for each swimming event, are well modelled by a GPD. This finding encourages us to consider the GPD as a marginal model for
all swimmers' available performances better than this same extreme threshold, which gives a suitable (negative) extreme threshold $u=-61.125$ seconds. This choice is further supported by asymptotic theory for univariate stationary processes that exhibit weak long-range dependence conditions, 
where the distribution of cluster maxima, and arbitrary values of the process excesses of a threshold, are identical in the limit as the threshold tends to the upper endpoint of the stationary distribution  \citep{leadbetter1991basis}.
The methods typically are based on the threshold-stability of the GPD, namely that if the GPD approximation \eqref{eq:probExc} is valid for exceedances above some threshold $u\in D_G$, then it holds for excesses over all higher thresholds $v$, where 
$v\in D_G$ and $v>u$. So if $u$ is the lowest threshold for which approximation~\eqref{eq:probExc} is exact, then any lower threshold 
will have excesses that do not follow
the GPD, whereas thresholds larger than $u$  ignore relevant observations and lead to inefficient inference.

Our model has two subject-specific parameters $\pmb{\theta}_i=(\alpha_i,\tau_i)$ per swimmer. Unless swimmer $i$ has undertaken sufficient swims in the data set then the posterior for the parameters for such swimmers will be weakly informed by the data, or even unidentifiable from the data if swimmer $i$ has only one recording. Here the standard Bayesian approach, and perhaps the most obvious, is to carry out analysis regardless and acknowledge that the marginal posterior distributions for such $\pmb{\theta}_i$ will be almost identical to the associated prior distributions. 
However, the prior on $\alpha_i$ is necessarily vague to allow for variation over swimmers, see Section~\ref{sec:swimPrior}, so the posterior information about these parameters adds little value to the overall inference. Moreover, it comes at a large computational cost from the many uninformative parameters, which requires the MCMC to do approximately twice as many of the slow 
 likelihood evaluations, see Section~\ref{sec:inference_issues}.
 Our analysis is instead restricted to only those swimmers with a ``sufficient'' number, i.e., more than $m$, of recordings in the data set. So, for the set of swimmers $\mathcal{I}_m$ that have recorded $m$ or fewer swims, i.e., for all $i\in \mathcal{I}$, with  $n_i:=|\mathcal{J}_i| \leq m$, these data are ignored.
The analysis is therefore conducted on the swimmers $\mathcal{I}\setminus \mathcal{I}_m$. A potential consequence of restricting the data set is that the GPD may no longer be a good fit to the tails of the data; however, we show in Section~\ref{sec: application} that this does not appear to be the case. Section~\ref{sec: discussion} discusses alternative approaches that do use the data for swimmers $\mathcal{I}_m$ and which do not suffer from computational complications, but they require additional modelling assumptions.  

If $m$ is chosen to be too small, some $\pmb{\theta}_i$ will have marginal posteriors with only minor differences from their priors and at the computational cost of needing more MCMC samples for convergence given the two additional variables per extra swimmer included. With $m$ too large, too much data are excluded so posteriors are less well informed than necessary. The strategy for choosing $m$ is to observe how the number of swimmers that have swum less than or equal to $m$, i.e., $k_m:=\sum_{i\in\mathcal{I}}\mathbbm{1}\{n_i \leq m\}$, varies with $m$. 
An abrupt increase was found when $m=7$.
Therefore, by selecting $m=7$, a relatively large proportion of those swimmers with only few observations $(40\%)$ are discarded, whilst only losing $20\%$ of the total observations. The final dataset used for analysis contained 120 swimmers, with 1435 total observations.
In an early analysis the model was fitted using only the 10 most prolific swimmers, i.e., $m=18$. Using these data the posterior means of the GPD parameters were very similar to those in the final analysis, reported in Section~\ref{results},
indicating that there is very little bias, or sensitivity, introduced through the choice of $m$.

\end{document}